\documentclass[preprint2]{aastex}
\usepackage{color}
\usepackage{amsmath}
\usepackage{natbib}
\usepackage{hyperref}
\usepackage{dcolumn}

\hypersetup{pdftitle = {Towards a Dynamical Collision Model},
pdfauthor = {Güttler et al.}, pdfstartview=FitH,
bookmarksopen=true, bookmarksnumbered=true}

\newcolumntype{.}{D{.}{.}{-1}}
\newcolumntype{d}[1]{D{.}{.}{#1}}

\shorttitle{Towards a Dynamical Collision Model}
\shortauthors{G\"uttler et al.}

\begin{document}

\title{The Physics of Protoplanetesimal Dust Agglomerates. IV. Towards a Dynamical Collision Model}%

\author{C. G\"uttler and M. Krause}
\affil{Institut f\"ur Geophysik und extraterrestrische Physik, Technische Universit\"at zu Braunschweig,
Mendelssohnstr. 3, D-38106 Braunschweig, Germany}%
\author{R. J. Geretshauser and R. Speith}
\affil{Institut f\"ur  Astronomie und  Astrophysik, Eberhardt Karls Universit\"at T\"ubingen,
Auf der Morgenstelle 10, D-72076 T\"ubingen, Germany}%
\and
\author{J. Blum}
\affil{Institut f\"ur Geophysik und extraterrestrische Physik, Technische Universit\"at zu Braunschweig,
Mendelssohnstr. 3, D-38106 Braunschweig, Germany}%

\begin{abstract}
Recent years have shown many advances in our knowledge of the collisional evolution of protoplanetary dust. Based on a variety of dust-collision
experiments in the laboratory, our view of the growth of dust aggregates in protoplanetary disks is now supported by a deeper understanding of
the physics involved in the interaction between dust agglomerates. However, the parameter space, which determines the collisional outcome, is
huge and sometimes inaccessible to laboratory experiments. Very large or fluffy dust aggregates and extremely low collision velocities are
beyond the boundary of today's laboratories. It is therefore desirable to augment our empirical knowledge of dust-collision physics with a
numerical method to treat arbitrary aggregate sizes, porosities and collision velocities. In this article, we implement
experimentally-determined material parameters of highly porous dust aggregates into a Smooth Particle Hydrodynamics (SPH) code, in particular an
omnidirectional compressive-strength and a tensile-strength relation. We also give a prescription of calibrating the SPH code with compression
and low-velocity impact experiments. In the process of calibration, we developed a dynamic compressive-strength relation and estimated a
relation for the shear strength. Finally, we defined and performed a series of benchmark tests and found the agreement between experimental
results and numerical simulations to be very satisfactory. SPH codes have been used in the past to study collisions at rather high velocities.
At the end of this work, we show examples of future applications in the low-velocity regime of collisional evolution.
\end{abstract}



\section{Introduction}
\subsection{\label{sec:growth}Protoplanetary dust growth}

The formation of planetesimals, the km-sized solid bodies whose further growth is controlled by mutual gravitational attraction, is still
enigmatic. Collisions among the dust aggregates are controlled by Brownian motion, drift motions with respect to the gas of the protoplanetary
disk, and turbulence in the gas \citep{Weidenschilling:1977a,WeidenschillingCuzzi:1993}. Once in contact, two dust grains experience a mutual
van der Waals force \citep{HeimEtal:1999}. From the theoretical and experimental standpoints, it is evident that the (sub-)micrometer-sized
protoplanetary dust grains initially undergo hit-and-stick collisions, which lead to the formation of fractal aggregates
\citep{WeidenschillingCuzzi:1993, BlumEtal:2000, KrauseBlum:2004}. As the collision energy increases, due to increasing aggregate mass and
collision velocity, dust aggregates undergo a restructuring phase, in which they acquire denser structures \citep{DominikTielens:1997,
BlumWurm:2000, WadaEtal:2007, WadaEtal:2008, WeidlingEtal:2009}. Laboratory experiments showed that collisions among the dust aggregates result
in fragmentation, i.e.\ in mass loss, if the impact velocities exceed $\sim 1 \, \rm m\,s^{-1}$ \citep{BlumWurm:2008}. Depending on the disk
model, this means that the direct collisional growth process ends (at the latest) at aggregate sizes for which this velocity is exceeded. For a
minimum-mass solar nebula model \citep{Weidenschilling:1977b, HayashiEtal:1985}, this size is approximately 10 cm.

The further growth is still highly speculative. \citet{WurmEtal:2001} and \citet{Blum:2004} proposed the accretion of collisional fragments by
aerodynamic and electrostatic effects, respectively. \citet{WurmEtal:2005} and \citet{TeiserWurm:2009} showed experimentally that a fraction of
a dust projectile can stick to a solidified larger dust target even at very large velocities. None of these processes, however, seem to work
globally and under all circumstances so that very specific conditions are required for the dust aggregates to grow at high impact velocities.
There is clearly a lack of understanding the detailed physics involved in the collisions between macroscopic dust aggregates of arbitrary
composition and porosity. Without better knowledge of the collisional physics of these bodies, any attempt to model the formation of
planetesimals as an aggregation process will have to fail.

\subsection{\label{sec:previous}Previous work}
In the three previous papers of this series, we described the collisional physics of high-porosity protoplanetary dust aggregates up to the
cm-size regime. In paper I \citep{BlumEtal:2006}, we introduced a method to experimentally produce monolithic dust aggregates with diameters of
2.5 cm. By choosing either monodisperse spherical monomer particles, quasi-monodisperse irregular particles, or polydisperse irregular grains,
we produced dust aggregates with volume filling factors (i.e.\ packing densities) $\phi = \rho / \rho_s$ of $\phi = 0.15$, $\phi = 0.11$, and
$\phi = 0.07$, respectively (see Table 1 in paper I for more details about the monomer-particle properties). Here, $\rho$ and $\rho_s$ are the
aggregate and the monomer density. Static uniaxial compression of these dust samples revealed that the maximum compaction for these
high-porosity dust aggregates is $\phi_{\rm max} = 0.20 \ldots 0.33$, a value very close to the overall porosity found in comets. The tensile
strengths of our dust samples were determined to $|T| = 200 \ldots 6,300$ Pa, depending on the monomer properties and the compaction. Also these
values are close to those found for comets. Paper II \citep{LangkowskiEtal:2008} concentrated on low-velocity impacts into these high-porosity
dust samples. We showed that sticking by penetration is the dominating process for impacts above a threshold velocity of $\sim 1 \, \rm
m\,s^{-1}$ for projectiles in the mm-size regime and flat dust targets. For shallow penetration, i.e.\ for impacts below the threshold velocity,
the projectiles bounce off, leaving a well-defined crater. It is obvious that the collisions result in the compaction of the target. In paper
III \citep{WeidlingEtal:2009}, we investigated the compaction for high-porosity mm-sized dust aggregates in bouncing collisions. Bouncing
collisions among dust aggregates show considerable energy losses \citep{BlumMuench:1993} so that it was natural to assume some degree of
compaction. In paper III, we found that -- although a single collision leads only to very localized compaction of the dust aggregate -- mm-sized
dust aggregates in protoplanetary disks can reach volume filling factors of $\phi \approx 0.35$ within a few dozen years.

\subsection{\label{sec:objectives}Objectives}

All previous experiments (see Sects.\ \ref{sec:growth} and \ref{sec:previous}) were limited by the experimentally available dust-aggregate sizes
and morphologies and the achievable collision velocities. In the astrophysical context, the need for numerical simulations of collisions between
dust aggregates of arbitrary composition, size and impact velocity arises from the fact that only a limited parameter space can be covered by
experiments. The ongoing debate about threshold velocities for sticking, bouncing, compaction, and fragmentation as well as the fragment size
distribution requires a thorough investigation of a wide range of collisions, varied over supposedly critical parameters, such as collision
velocity, porosity, size, impact parameter, impact angle and shape of the colliding dust aggregates. An extensive parameter study of that kind
is not feasible under laboratory conditions for the parameter ranges in question. Therefore, we aim to calibrate a Smooth Particle Hydrodynamics
(SPH) code and validate this model thoroughly with a series of independent benchmark tests. Hence, the SPH code gains a deeper reliability and
the conducted numerical simulations provide well-grounded insight into the physical behavior of dust aggregates.

\section{\label{sec:sph_basics}SPH in Dust Collisions}
SPH is a meshless Lagrangian particle method originally developed for astrophysical hydrodynamics applications. A detailed description of the
original SPH method may, e.g., be found in \citet{Monaghan:2005}.  The SPH code we utilize for the simulations in this work and the underlying
porosity model are introduced and described in full depth in \citet{GeretshauserEtal:preprint}. In the 1990s, SPH has been extended to model the
elastic and plastic behavior of solids, see e.g.\ \citet{LiberskyEtal:1993} and \citet{RandlesLibersky:1996}. The continuous solid objects are
discretized into interacting mass packages called particles, which form a natural frame of reference for any deformation and fragmentation that
may occur.

The SPH code solves the equations of continuum mechanics in Lagrangian form, in particular the continuity equation
\begin{equation}
    \frac{\mathrm{d}\rho}{\mathrm{d}t} + \rho \frac{\partial v_\alpha}{\partial x_\alpha} = 0,
\end{equation}
and the equation of motion
\begin{equation}
    \frac{\mathrm{d} v_\alpha}{\mathrm{d}t} = \frac{1}{\rho} \frac{\partial \sigma_{\alpha\beta}}{\partial x_\beta}.
\end{equation}
Here, Einstein's summing convention holds throughout and Greek indices denote spatial coordinates. The variables have their usual meanings,
i.e., $\rho$ denotes the density, $v$ the velocity, and $\sigma_{\alpha\beta}$ the stress tensor. The latter is defined according to
\begin{equation}
    \sigma_{\alpha\beta} = -p \delta_{\alpha\beta} + S_{\alpha\beta},
\end{equation}
consisting of a pressure part with pressure $p$ and and a shear part given by the traceless deviatoric stress tensor $S_{\alpha\beta}$.

The deviatoric stress is defined by the constitutive equations. To model elastic behavior according to Hooke's law we adopt the approach by
\citet{Benz:1994} for the time evolution of the deviatoric stress,
\begin{equation}
    \frac{\mathrm{d} S_{\alpha\beta}}{\mathrm{d} t} =
    2\mu \left( {\dot{\epsilon}}_{\alpha\beta} - \frac{1}{d} \delta_{\alpha\beta} {\dot{\epsilon}}_{\gamma\gamma}\right)
    + S_{\alpha\gamma}R_{\gamma\beta} + S_{\beta\gamma}R_{\gamma\alpha},
\end{equation}
where $\mu$ is the shear modulus and $d$ denotes the dimension. The rotation rate tensor $R_{\alpha\beta}$ reads
\begin{equation}
    R_{\alpha\beta} =
    \frac{1}{2} \left( \frac{\partial v_\alpha}{\partial x_\beta} - \frac{\partial v_{\beta}}{\partial x_\alpha} \right)
\end{equation}
and the strain rate tensor ${\dot{\epsilon}}_{\alpha\beta}$ accordingly
\begin{equation}
    {\dot{\epsilon}}_{\alpha\beta} = \frac{1}{2} \left( \frac{\partial v_\alpha}{\partial x_\beta} + \frac{\partial v_{\beta}}{\partial x_\alpha} \right).
\end{equation}
This set of equations is closed by a suitable equation of state and describes the elastic behavior of a solid body.

\begin{figure}[t]
    \center
    \includegraphics[width=7.5cm]{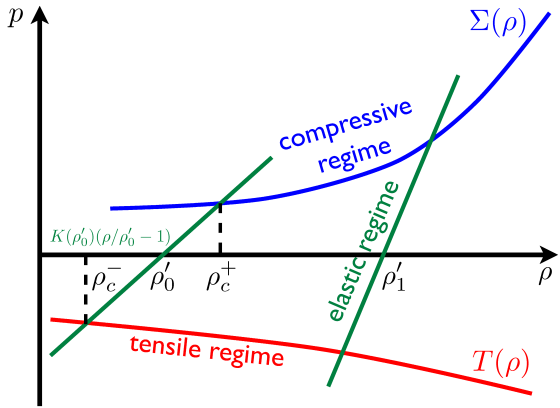}
    \caption{\label{fig:sirono-model}The modified Sirono porosity model is divided into the elastic, compressive and tensile regimes. The latter two
    are reached by exceeding the compressive and tensile strength, respectively, which leads to irreversible plastic deformation.}
\end{figure}

In order to simulate also the plastic behavior of porous bodies, we adopt a modified version of the porosity model by \citet{Sirono:2004}
(Fig.~\ref{fig:sirono-model}). According to this approach, plasticity is modeled within the equation of state, which is divided into three
different regimes. In the first regime, plastic behavior is caused by compression that exceeds a critical limit, the compressive strength
$\Sigma(\rho)$, while in the second regime, tension exceeds the tensile strength limit $T(\rho)$. In between these limits, the third, the
elastic regime of the material is described by a special version of the Murnaghan equation of state. Thus, the full equation of state reads
\begin{equation}
    p(\rho) = \left\{
    \begin{array}{lc}
    \Sigma(\rho) & \rho > \rho_{\rm c}^+ \\
    K({\rho'}_0) (\rho/{\rho'}_0 - 1) & \rho_{\rm c}^- \le \rho \le \rho_{\rm c}^+ \\
    T(\rho) & \rho < \rho_{\rm c}^- \\
    \end{array}
    \right.\;.\label{eq:pressure_overview}
\end{equation}
The quantity ${\rho'}_0$ denotes the reference density, which is the density of the material without any external stress. $\rho_{\rm c}^+$ and
$\rho_{\rm c}^-$ are limiting quantities, where the transition between the elastic and plastic regime for compression and tension, respectively,
takes place. Once these limits are exceeded, the material leaves the elastic path which represents the path where energy is conserved, and loses
internal energy by following the paths of the compressive and tensile strength (Fig.~\ref{fig:sirono-model}).

\section{Towards an Equation of State for Dust Aggregates}

In this laboratory section, we will provide the macroscopic material parameters, which are necessary for the SPH model introduced in section
\ref{sec:sph_basics}. We recapitulate the tensile strength measurements of \citet{BlumSchraepler:2004} and give an interpolation for different
volume filling factors. The compressive strength for unidirectional (1D) compression was also measured by \citet{BlumSchraepler:2004}, while in
this paper we will present measurements on omnidirectional (3D) compression. Moreover, we will introduce a simple impact experiment, which will
be used for calibrating the SPH model: A mm-sized glass bead (or a glass bead analog) impacts into a well-defined 2.5~cm dust sample at a
collision speed of 0.1 to 1~m~s$^{-1}$. The dust sample, consisting of 1.5~$\mu$m SiO$_2$ monodisperse spheres, was formed by random ballistic
deposition (RBD) and has therefore a volume filling factor of $\phi_0=0.15$ (see \citet{BlumSchraepler:2004} and references therein). The
deceleration curve, penetration depth and impact duration of the glass bead are measured as well as the compression of the dust beneath the
glass bead to compare these results with an impact computed by the SPH model.

\subsection{\label{sec:tensile}Tensile Strength}

In \citet{BlumSchraepler:2004} and paper I, we reported on measurements of the tensile strength of dust samples of various constitutions (i.e.\
monomer size distribution, morphology, composition, volume filling factor). The best set of data was collected for the dust aggregates
consisting of spherical 1.5~$\mu$m SiO$_2$ monomers (see above). For packing densities of $\phi = 0.15$, $\phi = 0.41$, $\phi = 0.54$ and $\phi
= 0.66$, we found tensile strengths of $|T| = 1,000$ Pa, $|T| = 2,400$ Pa, $|T| = 3,700$ Pa and $|T| = 6,300$ Pa, respectively. To a good
approximation, these values can be expressed by a relation of the form
\begin{equation}
    T(\phi) = - \left( 10^{2.8 + 1.48 \, \phi} \right) \label{eq:tensile_strength} \;.
\end{equation}
This expression will be used throughout this paper for the packing-density-dependence of the tensile strength.

\subsection{Static Measurement of Compressive Strength Curves}
\label{sec:compr_curve}

The compression curve of a given material tells us how the material behaves under an applied pressure $\Sigma$ in changing its volume filling
factor $\phi$. If the material can be described by macroscopic parameters, the volume filling factor is representative for the material density
and so the development of the compression curve $\phi(\Sigma)$ (cf. Eq.\ \ref{eq:pressure_overview}) is essential to establish a collision model
and learn about collisions of protoplanetary dust aggregates.

Measurements of the compression curve were already performed by \citet{BlumSchraepler:2004} and in paper I. We will again focus on the dust
samples made of 1.5~$\mu$m SiO$_2$ spheres, whose properties are compiled in Table 1 in \citet{BlumSchraepler:2004}. In the compression
experiments of \citet{BlumSchraepler:2004}, a dust sample was fixed between two parallel glass plates, which were then pushed together with an
increasing force. The measurement of the dust mass, dust volume, compression force, and, thus, pressure yield the compression curve
$\phi(\Sigma)$. The force was applied in one direction, which is therefore called unidirectional compression. The dust sample flattens in the
direction of the force but, at the same time, also expands in the other directions. For dust samples made of 1.5~$\mu$m SiO$_2$ spheres, this
leads to an equilibrium filling factor of 0.33 for pressures exceeding $10^5$~Pa. This compression curve is only applicable to protoplanetary
dust collisions, if the material compressed in the impact zone creeps sideways as it did in the static experiments. As we will show later by
x-ray analysis of the compression next to an impact site, this is not the case.

\begin{figure}
    \center
    \includegraphics[width=7.5cm]{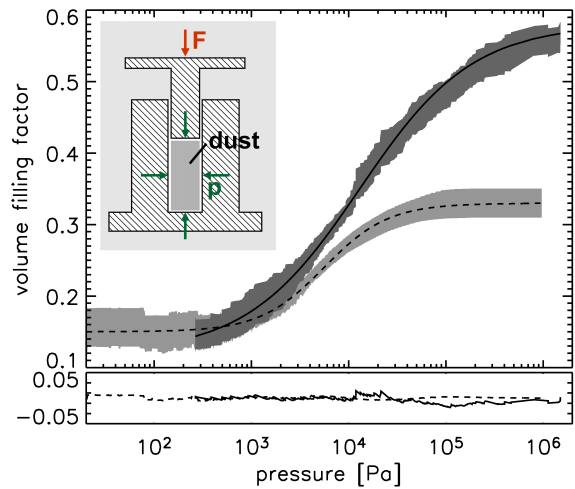}
    \caption{\label{fig:omni_compr_curve}The omnidirectional (solid line) and unidirectional (dashed line) static compression curves and the
    standard deviations of the measurements (gray shaded areas). The upper plot shows the analytical approximations from Eq.\ \ref{eq:approx_phi} and
    Table \ref{tab-compr_curve_param}, while the lower plot gives the deviation between approximation and measurement. The inset describes the setup
    for the omnidirectional measurement.}
\end{figure}

Consequently, a second way to measure the compressive strength curve is to fix the dust sample at the sides with closed walls. In this case, the
pressure cannot be released and acts from all sides, thus omnidirectional compression. We performed experiments in which we cut a cylindrical
section from an RBD dust sample with a thin-walled plastic tube of 7~mm diameter. This cylindrical dust sample of approximately 1~cm height was
then put into a 7~mm borehole in an aluminum block. Carefully pushing a piston into this borehole leads to an omnidirectional pressure onto the
dust sample (see inset in Fig.~\ref{fig:omni_compr_curve}). The setup was put onto a balance and the piston was loaded with weights of
increasing mass. This weight force, divided by the piston area, yields the pressure $\Sigma$, while the mass and height of the dust sample
determine the volume filling factor $\phi$ (Fig.~\ref{fig:omni_compr_curve}). Due to the fact that the dust sample is not a frictionless fluid,
force chains inside the sample might locally reduce the pressure. Thus, the pressure for the idealized compression curve can be slightly lower.

The solid line in Fig.~\ref{fig:omni_compr_curve} denotes an analytical approximation of the mean filling factor of nine individual experiments
as a function of the applied pressure and the gray shaded area is the standard deviation of the measurements. The analytical function is based
on a Fermi distribution with logarithmic pressure in the energy term
\begin{equation}
    \phi(\Sigma)=\phi_2-\frac{\phi_2-\phi_1}{\exp\left(\frac{\lg\Sigma-\lg p_{\rm m}}{\Delta}\right)+1}\label{eq:approx_phi}
\end{equation}
and is only valid for $\phi\geq\phi_0$. For pressures below $\Sigma(\phi_0)$ the dust aggregate behaves elastically. The parameters for the
unidirectional and omnidirectional compression curve are given in Table \ref{tab-compr_curve_param}. The bottom plot in
Fig.~\ref{fig:omni_compr_curve} gives the deviation between the analytical approximation and the data, which is within $\phi_{\rm err}=\pm0.02$.
Often, the inverse function $\Sigma(\phi)$ is used (see Eq.\ \ref{eq:pressure_overview}), which is here
\begin{equation}
    \Sigma(\phi)=p_{\rm m}\cdot \left(\frac{\phi_2-\phi_1}{\phi_2-\phi}-1\right)^{\Delta\cdot\ln 10}\;.\label{eq:strength_curve}
\end{equation}

\begin{table}[t]
    \caption{\label{tab-compr_curve_param}Parameters of the analytical approximation for the two compression curves}
    \begin{tabular}{lccd{2.1}c}
    \hline
    & $\phi_1$ & $\phi_2$ & \multicolumn{1}{c}{$p_{\rm m}$ [kPa]} & $\Delta$ [dex]\\
    \hline
    unidirectional  & 0.15 & 0.33 &  $5.6$ & 0.33 \\
    omnidirectional & 0.12 & 0.58 & $13.0$ & 0.58 \\
    \end{tabular}
\end{table}

Compared with the unidirectional compression curve (dashed line) of \citet{BlumSchraepler:2004}, the filling factor also starts off at the
original dust sample filling factor of $\phi_0=0.15$ \citep[cf.][]{BlumSchraepler:2004}, but diverges from the unidirectional curve for
pressures $p\gtrsim10^3$~Pa. For those pressures, the filling factor is systematically higher, meaning on the other hand that the same filling
factor is much easier to achieve if the pressure acts from all sides. So far there was no equilibrium filling factor found like in the case of
the unidirectional compression experiments. The filling factor still significantly increases for the highest applied pressure of $10^6$~Pa.
However, the analytical approximation indicates an equilibrium for $\phi_2=0.58$, which is not far from  random close packing of monodisperse
spheres \citep[$\phi\simeq0.64$, see e.g.][]{TorquatoEtal:2000}, the maximal possible compression without breaking the dust grains.

The new compression curve is still a static measurement. It is applicable for omnidirectional static pressures like the hydrostatic equilibrium
inside planetesimal bodies. It is questionable if this compression curve is valid for dynamic collisions but it is a second attempt to assume
that surrounding material, which does not interact in a collision, acts as a confining wall to the active impact volume instead of creeping
sideways.

\subsection{Deceleration Experiments}

\subsubsection{Experimental Setup}

\begin{figure}[t]
    \center
    \includegraphics[width=7cm]{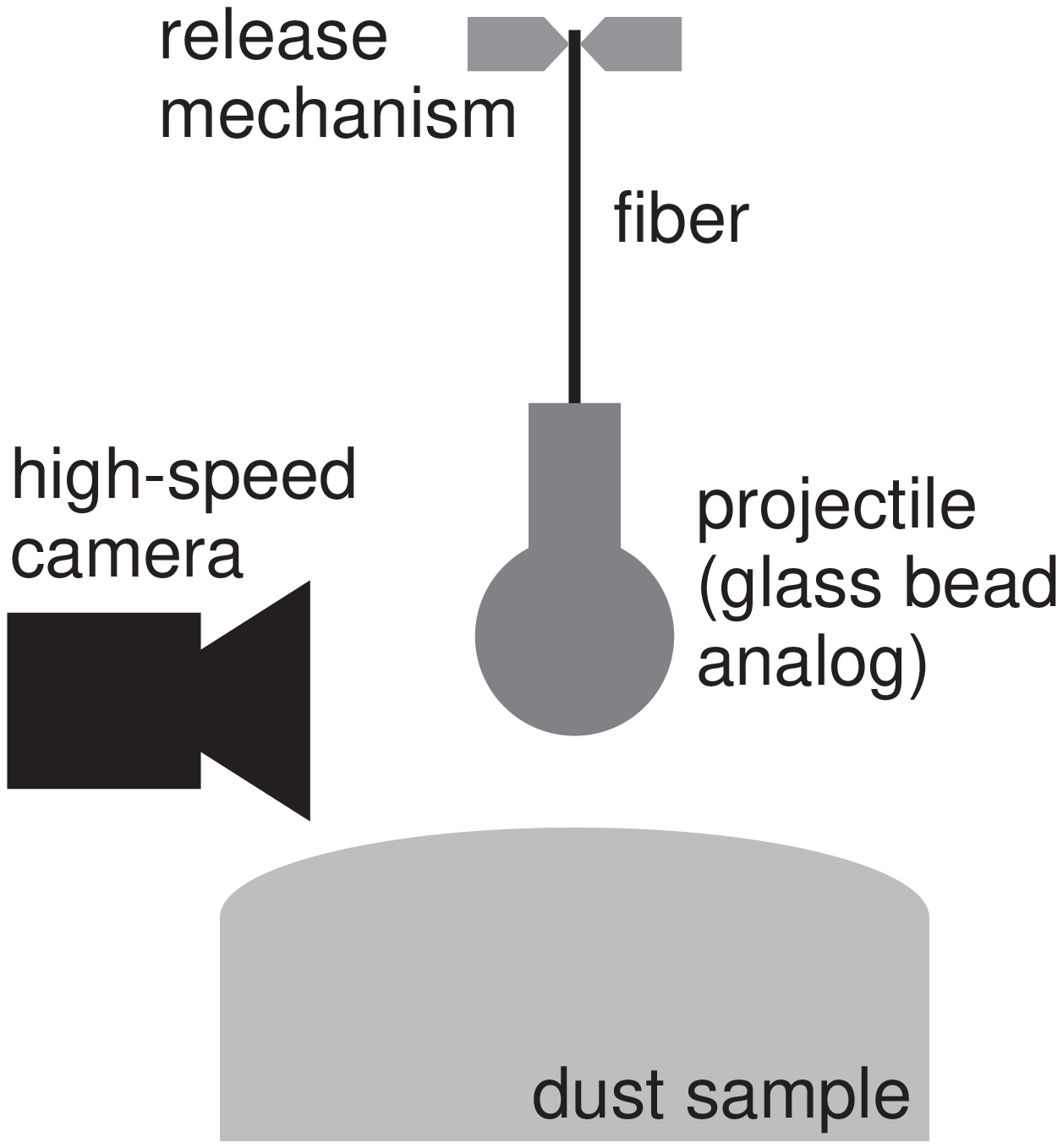}
    \caption{\label{fig:deceleration_sketch}Setup for the deceleration measurement: An elongated projectile as a glass bead analog was dropped into
    the dust sample. Before dropping from a height of 1 to 40~mm, it was suspended on a fiber  with negligible mass  to
    avoid rotational motion. A high-speed camera observes the deceleration of the projectile.}
    \includegraphics[width=7.5cm]{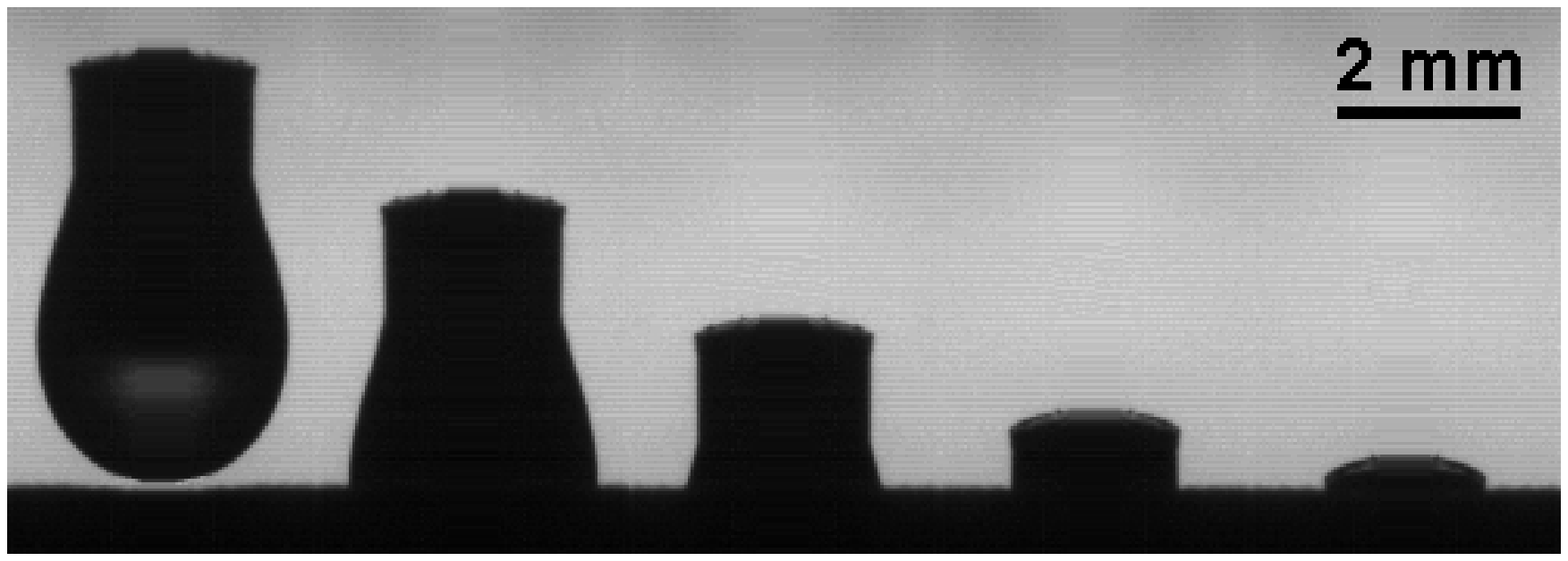}
    \caption{\label{fig:deceleration_sequence}Image sequence of a decelerated projectile. The time between two images is 1.7~ms.}
\end{figure}

The experimental setup consists of a vacuum chamber (gas pressure $\sim0.5$~mbar) in which a projectile is suspended on a thin fiber
(Fig.~\ref{fig:deceleration_sketch}) with negligible mass to prevent rotation and lateral velocities.
The distance between the suspended projectile and the surface of the dust sample determines the impact velocity $v_0$. The projectile consists
of an elongated solidified epoxy droplet at the bottom and a cylindrical plastic tube at the top end. After the release of the projectile, it is
accelerated by gravity and decelerated once it is in contact with the dust sample. The deceleration within the dust sample is observed by a
high-speed camera (Fig.~\ref{fig:deceleration_sequence}). From the deceleration curve of the projectile we can derive fundamental dynamic
properties of the target dust aggregate.

The bottom shapes of the projectiles were spherical with diameters of $R \approx 0.5~$mm and $R \approx 1.5$~mm and masses of $m\approx1~$mg and
$m\approx30~$mg, respectively (see Table \ref{tab-dec_results}). The effective densities of the projectiles of $\rho =\rm ~2,400 \ldots3,100
~kg~m^{-3}$ match those of the astronomically relevant silicates, while the combination of low-density epoxy and plastic tube made sure that the
top of the projectile was always visible to the camera even if the intrusion depth was larger than the projectile diameter. The high-speed
camera was operated at a frame rate of 12,000 frames per second with a resolution of $\sim30\;\mu$m/pixel and the position of the upper edge of
the projectile was measured with sub-pixel accuracy of $\rm \sim 3 ~\mu m$. The first touch of the projectile with the surface of the dust
sample marks the time $t=0$ and can clearly be determined from the deviation of the trajectory compared to a free falling projectile. After its
deepest penetration, the projectile bounces back (by $\sim100\;\mu$m) and oscillates in the vertical direction, which we will not take into
account in the further discussion.

\subsubsection{\label{sec:dec_curve_results}Experimental Results}

\begin{table*}[t]
    \scriptsize
    \caption{\label{tab-dec_results}Experimental results of the deceleration experiments.}
    \begin{tabular}{ccd{2.1}ccccd{2.2}d{2.2}}
        \hline
        experi- & projectile      & \multicolumn{1}{c}{projectile} & effective          & impact      & penetration & stopping   & \multicolumn{1}{c}{standard}               & \multicolumn{1}{c}{standard} \\
        ment    & diameter        & \multicolumn{1}{c}{mass}       & projectile         & velocity    & depth       & time       & \multicolumn{1}{c}{deviation $\sigma$ for} & \multicolumn{1}{c}{deviation $\sigma$ for} \\
        number  & $2\cdot R$ [mm] & \multicolumn{1}{c}{$m$ [mg]}   & density [kg/m$^3$] & $v_0$ [m~s$^{-1}$] & $D$ [mm]    & $T$ [ms]   & \multicolumn{1}{c}{polynomial [$\mu$m]}    & \multicolumn{1}{c}{sine [$\mu$m]} \\ \hline
        1&2.73&25.7&2412&0.89&3.16&5.92& 2.43&21.48\\
        2&2.94&32.0&2404&0.85&3.08&5.93& 2.99&22.64\\
        3&2.77&26.4&2372&0.73&2.70&6.09& 1.87&17.29\\
        4&2.94&32.0&2404&0.17&0.70&6.18& 4.63&10.78\\
        5&2.94&32.0&2404&0.16&0.75&7.13& 2.84& 7.69\\
        6&2.94&32.0&2404&0.20&0.80&5.92& 2.96& 8.95\\
        7&2.94&32.0&2404&0.32&1.31&6.47& 1.62& 2.53\\
        8&2.77&27.6&2480&0.50&2.03&6.38& 1.82& 3.57\\
        9&2.77&26.4&2372&0.37&1.33&5.71& 3.65& 6.26\\
        10&0.99& 1.5&2854&0.67&1.06&3.09& 3.48&27.77\\
        11&0.99& 1.5&2854&0.76&1.15&2.94& 1.41&23.24\\
        12&0.99& 1.5&2854&0.79&1.47&3.12& 14.29& 5.62\\
        13&0.85& 1.0&3109&0.19&0.32&2.46& 1.94& 3.79\\
        14&0.85& 1.0&3109&0.35&0.83&3.28& 2.60&11.15\\
        15&0.85& 1.0&3109&0.36&0.72&3.11& 2.71& 3.73\\
        \hline
    \end{tabular}
\end{table*}

We performed 15 impacts of our projectiles into the porous dust samples, which are compiled in Table \ref{tab-dec_results}. The time-resolved
deceleration data $h(t)$ were cleaned from gravitational influence by adding $\frac{1}{2}gt^2$ to the negative intrusions so that the
gravity-independent deepest penetration depth $D$ and stopping time $T$ could be determined. The intrusion curves were normalized in space and
time through $h'(t') = h(t) / D$ and $t' = t / T$ so that $h'(t'=0)=0$ (first contact) and $h'(t'=1)=-1$ (deepest intrusion), and can then be
well represented by a sine function
\begin{equation}
    h'(t')=-\sin\left(t'\cdot\frac{\pi}{2}\right)\;.
\end{equation}
Alternatively, a fourth order polynomial with only one free parameter was used for fitting the data, where the mean standard deviation between
the fit and the $N$ data points $\sigma=\sqrt{\frac{1}{N}\sum_{i=1}^N(h'(t_i')-h_i')^2}$ amounts to only 2 -- 4~$\mu$m in absolute units (cf.
Table \ref{tab-dec_results}).
\begin{figure}
    \center
    \includegraphics[width=7.5cm]{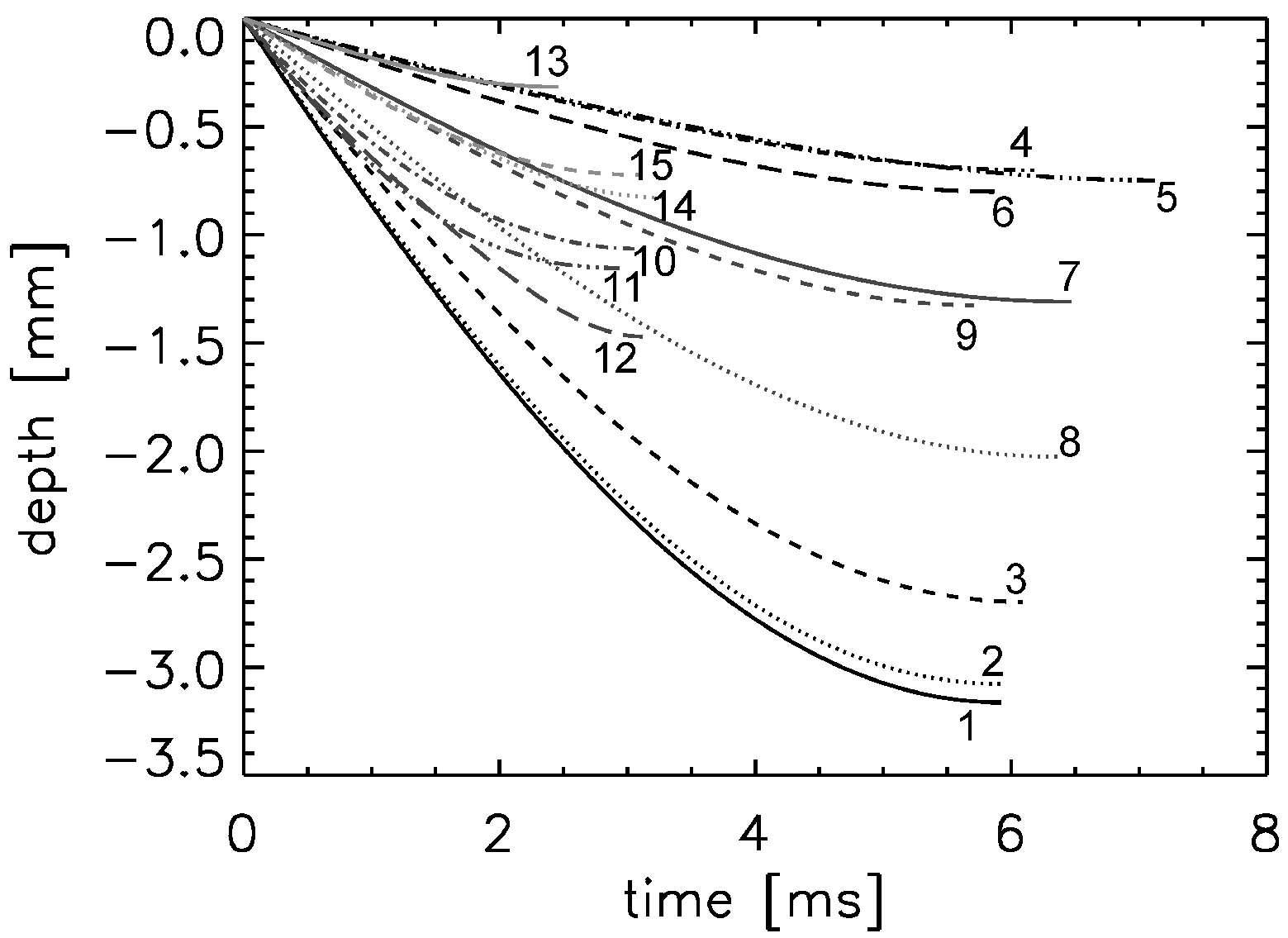}
    \caption{\label{fig:abs_dec_curve}Fitted deceleration curves in absolute units. The experiment numbers at the deceleration curves confer to
    those in Table \ref{tab-dec_results}.}
\end{figure}
Although the standard deviation for the sine function is rather of the order of 10~$\mu$m, we take the sine
function because it has no free parameter and the standard deviation is still less than the pixel size of 30~$\mu$m. In few experiments, the
1~mm projectiles canted over before coming to rest. In these cases, the data was used as long as reliable and the remaining deceleration curve
and, thus, the penetration depth was extrapolated.

Figure \ref{fig:abs_dec_curve} shows all measured deceleration curves in absolute units. Different intrusion depths and stopping times can
clearly be distinguished in this plot. The intrusion depths increase with increasing impact velocities (i.e.\ with the absolute values of the
initial slopes of the curves), while the stopping times are rather constant for one projectile size ($T \sim 6$~ms for 3~mm projectiles
[nos.~1-9 in Fig.~\ref{fig:abs_dec_curve}] and $T \sim 3$~ms for 1~mm projectiles [nos.~10-15 in Fig.~\ref{fig:abs_dec_curve}]) and, thus,
independent from the impact velocity $v_0$.
\begin{figure}[t]
    \center
    \includegraphics[width=7.5cm]{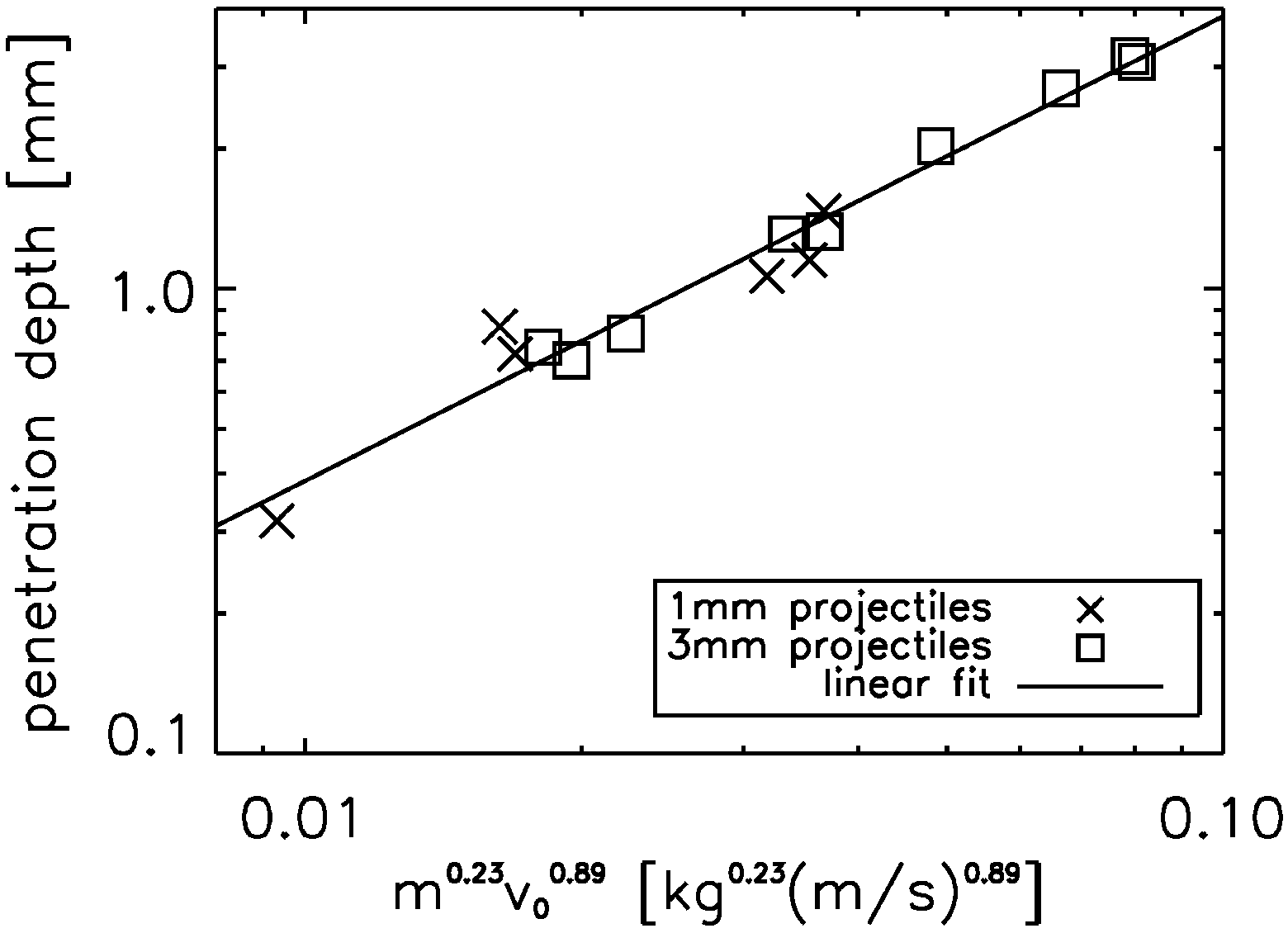}
    \caption{\label{fig:penetration_depth}The best relation for the penetration depths from a $\chi^2$ test yields a dependence of $D\propto
    m^{0.23}v_0^{0.89}$. The intuitive relation $D\propto mvA^{-1}$ is possible within the uncertainties.}
    \includegraphics[width=7.5cm]{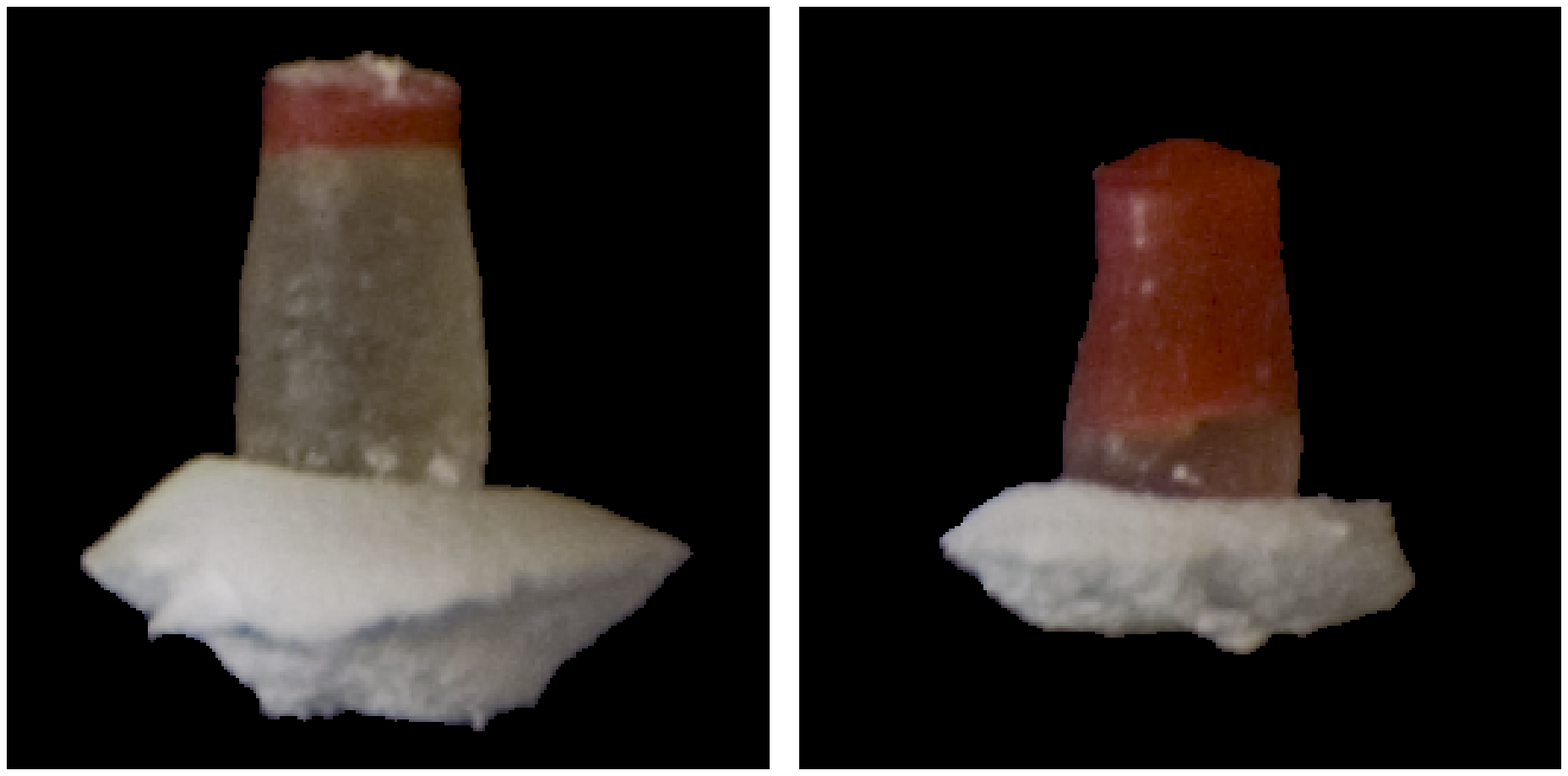}
    \caption{\label{fig:dusty_projectiles}Dust sticks to the projectiles after pulling them out of the dust sample. This is an indication of
    compacted material under the projectile as will be confirmed in Sect.\ \ref{sec:dyn_compr_exp}.}
\end{figure}
A $\chi^2$ test yielded the best-fitting power-law relation between the penetration depth, impact velocity and mass of the form
\begin{equation}
    D=\gamma_\mathrm{D} \cdot m^{\alpha_\mathrm{D}} \cdot v_0^{\beta_\mathrm{D}}\;,\label{eq:penetr_depth}
\end{equation}
with $\alpha_\mathrm{D} = 0.23 \pm 0.13$, $\beta_\mathrm{D} = 0.89 \pm 0.34$, and $\gamma_\mathrm{D} =(3.86 \pm 0.11) \cdot 10^{-2} ~\rm
kg^{-0.23} m^{0.11} s^{-0.89}$ (Fig.~\ref{fig:penetration_depth}). The respective errors denote the $1\sigma$ uncertainties.  A more intuitive
relation would be $D \propto mv_0A^{-1}$, with $A=\pi R^2$ being the cross section of the projectile.  This relation has a clear physical
meaning as the penetration depth is determined by the quotient of the momentum $mv_0$ as driving force and the cross sectional area $A$ as
resistive parameter. With $\alpha_\mathrm{D}=\frac{1}{3}$ and $\beta_\mathrm{D} = 1$ being possible within the uncertainties, the linear
relation $D \propto mv_0A^{-1} \propto \rho_\mathrm{p}^{1/3}Rv_0$ is also possible. However, constraining the exponent $\delta_\mathrm{D}$ as $D
\propto \rho_\mathrm{p}^{\delta_\mathrm{D}}$ in Eq. \ref{eq:penetr_depth} was unfortunately not feasible due to the too small variations in the
effective projectile density $\rho_\mathrm{p}$ (cf. Table \ref{tab-dec_results}).

For the stopping time, we found
\begin{equation}
    T=\gamma_T\cdot m^{\alpha_T} \cdot v_0^{\beta_T} \label{eq:stopping_time}
\end{equation}
with $\alpha_T = 0.23 \pm 0.08$, $\beta_T = 0.01 \pm 0.23$, and $\gamma_T = (6.77 \pm 0.20) \cdot 10^{-2} {\;\rm
kg^{-0.23} m^{-0.01} s^{1.01}}$ was found for the stopping time.

After pulling the projectiles out of the dust sample, dust stuck to the surface with which it had been in contact before
(Fig.~\ref{fig:dusty_projectiles}). With the preliminary assumption that this is compacted dust and the layer where it broke off is the
transition from compacted to non-compacted dust (transition in tensile-strength), this gives an indication for the compressed volume which will
be analyzed in detail in the forthcoming section.

\subsection{\label{sec:dyn_compr_exp}Dynamic Compression Experiments}

\subsubsection{Experimental Setup}
\begin{figure}[b]
    \center
    \includegraphics[width=7cm]{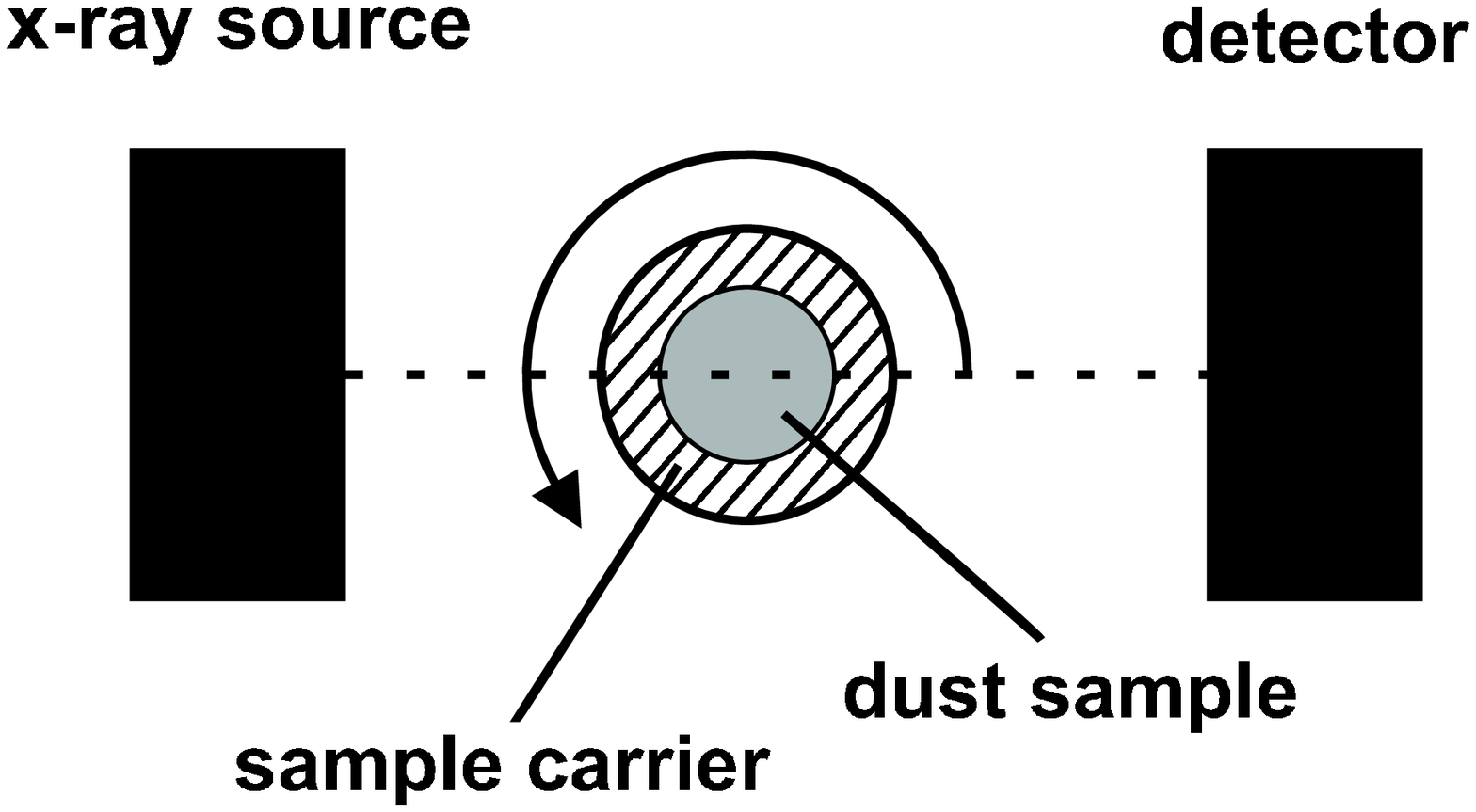}
    \caption{\label{fig:xrt_setup}Setup of the x-ray micro-CT measurement: the sample is rotated between an x-ray source and a detector. A 3D
    density reconstruction can be computed from the transmission images.}
\end{figure}
In order to investigate in more detail the compression behavior of the dust aggregates by collisions, we performed impact experiments with
subsequent x-ray micro-tomography measurements to analyze the degree of compaction.

Under vacuum conditions we dropped a single glass spherule with a diameter of $\sim$ 1~mm from a given height of $\sim$ 75 mm into an RBD dust
sample within a plastic tube with 7~mm diameter. To ensure the sphere to preferably hit the center of the dust sample within the narrow plastic
tube, the released projectile was guided by falling through a tube. Due to friction and collisions with the tube's walls the impact velocity of
(0.8$\pm$0.1)~m~s$^{-1}$, that was independently measured by high-speed imaging in 10 drops, is much lower than expected from free fall.
However, the velocity in the specific experiment was not measured and can well be in the lower range of the error. From comparison of the
observed penetration depth (see Fig.~\ref{fig:color_contour} in Sect. \ref{sec:density_plots}) with the results in
Fig.~\ref{fig:penetration_depth} we expect a velocity of $v = 0.65$~m~s$^{-1}$, which we will use for the further study.

For analyzing the density distribution of the dust sample cutout with the embedded glass sphere, the dust sample was scanned by an x-ray
micro-computer-tomograph (Micro-CT SkyScan 1074) at the University of Osnabr\"uck. The dust sample was positioned on top of a rotatable sample
carrier between the x-ray source and the detector (CCD camera) (Fig.~\ref{fig:xrt_setup}). While rotating stepwise around by $360^{\circ}$, 400
transmission images were captured. Based on this data set, a 3-dimensional density reconstruction was calculated by the SkyScan Cone-Beam
Reconstruction Software provided with the x-ray micro-CT instrument.

\subsubsection{\label{sec:density_plots}Experimental Results}

\begin{figure}[t]
    \center
    \includegraphics[width=5cm,angle=90]{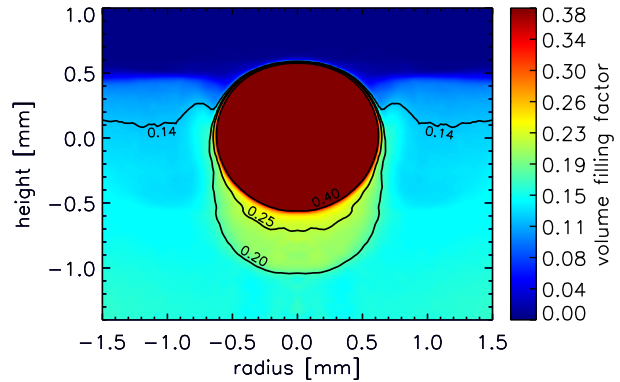}
    \caption{\label{fig:color_contour}Spatial averaged volume filling factors mirrored at the vertical center line. The volume under the sphere is
    compacted to a volume filling factor $>0.2$ (yellow), while the surrounding material is nearly unaffected (light blue).}
\end{figure}

In the following we present the results of two impact experiments. Further experiments with differently sized spheres and different impact
velocities are intended. To visualize the spatial density distribution of the observed dust sample with the impacted glass sphere, the
3-dimensional reconstruction data was cylinder-symmetrically averaged with the vertical axis aligned with the sphere center. Figure
\ref{fig:color_contour} displays the mean volume filling factor as a function of height and radius, whereas the data is mirrored with respect to
the vertical center line of the diagram.
The color gradient from yellow to light blue underneath the impacted sphere (red color: saturated density values of the considerably denser
glass spherule) clearly shows the densification of the porous dust sample with an initial volume filling factor of $\phi_0\approx$ 0.15. The
compressed area, emphasized by overplotted contour-lines, is located almost cylindrically shaped beneath the sphere and extends only slightly to
the lateral borders of the sphere. Thus, the assumption of an omnidirectional compression curve, made in Sect.\ \ref{sec:compr_curve}, seems to
be justified.

Analysis of the distribution of occurring volume filling factors related to their fraction of volume within an uncompressed dust sample
\begin{figure}[t]
    \center
    \includegraphics[width=7.5cm]{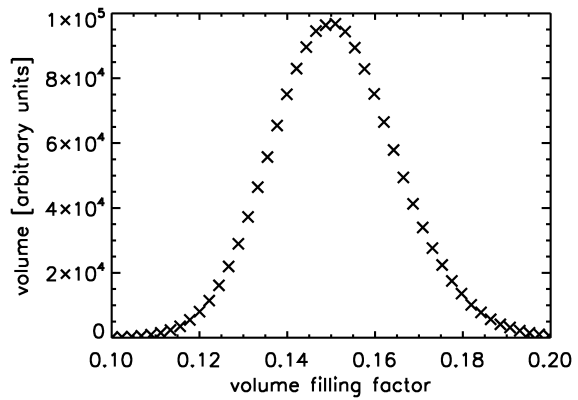}
    \caption{\label{fig:cake_density_histogram}Distribution of volume filling factors for an uncompressed dust sample, which follows a Gaussian
    distribution with a mean value of $\phi \approx$ 0.15.}
\end{figure}
\begin{figure}[t]
    \center
    \includegraphics[width=7.5cm]{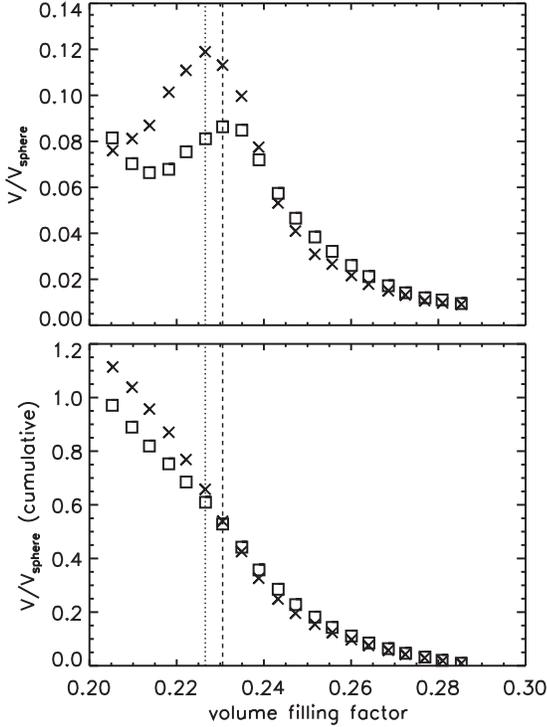}
    \caption{\label{fig:vol_fillfactor}\textbf{Top:} Distribution of volume filling factors only for the compressed area underneath the impacted
    sphere for two experiments. The dashed and dotted lines mark the most occurring volume filling factors for each curve, lying at $\phi \approx$
    0.23. \textbf{Bottom:} Normalized volume fraction of compacted area corresponding to a volume filling factor $> \phi$.}
\end{figure}
provides a Gaussian-shaped distribution with a mean value of $\phi \approx$~0.15 (Fig. \ref{fig:cake_density_histogram}). Figure
\ref{fig:vol_fillfactor} (top) shows the volume fraction (normalized by the sphere volume) of volume filling factors, which we determined only
regarding the compacted  volume  underneath the impacted sphere for the two impact experiments. In both curves the most prominent volume filling
factor is around $\phi=0.23$, indicated by the dashed and dotted lines. The decreasing left flank of the curves corresponds to the transition
region between compressed and uncompressed dust material (see right curve flank of Fig.~\ref{fig:cake_density_histogram}). The same data plotted
in a cumulative way (Fig. \ref{fig:vol_fillfactor}, bottom), represent the amount of compacted volume in units of the sphere volume that
complies with a volume filling factor greater than a certain value. According to the volume filling factor values at the boundary to the
uncompressed dust, given by the minima of the left side of the curves in Fig. \ref{fig:vol_fillfactor} (top), we can conclude from the
cumulative curves (Fig.~\ref{fig:vol_fillfactor}, bottom) that the compressed  volume  due to an impacting sphere of 1~mm size into a
high-porosity dust sample ($\phi \approx$~0.15) fills the volume of $\sim$ 0.8-1.2~sphere volumes.

\subsection{\label{sec:shift_strength_curve}Requirements of a Dynamic Compressive Strength Curve}
As seen in the previous sections, we have abundant indirect information about the compression behavior of loose dust samples. However, the basic
question how the dynamic compressive strength curve, $\phi(\Sigma)$, looks, remains unanswered. We approach this problem the following way: (1)
For low compressions, $\Sigma \rightarrow 0$, the volume filling factor is given by the initial properties of the material, i.e.\ $\phi
\rightarrow \phi_1$ (see Table \ref{tab-compr_curve_param}). (2) The maximum compression for $\Sigma \rightarrow \infty$ is given by the value
$\phi_2$ in Table \ref{tab-compr_curve_param} for the omnidirectional case, because the XRT analysis shows no material creeping sideways as was
the case for the unidirectional flow \citep[see][]{BlumSchraepler:2004}. (3) With these two limits in mind, we apply Eq.\ \ref{eq:approx_phi} as
an approximation to the functionality of the dynamic compressive strength, which leaves us with the two free parameters $\Delta$ and $p_{\rm
m}$. The maximum slope of the compression at $\Sigma = p_{\rm m}$ is given by ${\rm d}\phi/{\rm d} \lg\Sigma = (\phi_2 - \phi_1)/\Delta$. For
the unidirectional and omnidirectional static curves, we get slope values of 0.55 and 0.79, respectively (see Table
\ref{tab-compr_curve_param}). These are in fact not so different so that we adopt for the dynamic case the slope of the omnidirectional
compression. Thus, we assume $\Delta = 0.58$ dex for the dynamic case.  A refined study that takes both, $\Delta$ and $p_{\rm m}$, as free
parameters will be conducted in \citet{GeretshauserEtal:preprint}, but in this paper we only vary $p_{\rm m}$.


\section{Calibrating the SPH code}

The laboratory experiments in the previous section provided the static omnidirectional compressive strength $\Sigma$ and the tensile strength
relation $T$ as most important ingredients for the Sirono porosity model implemented in the SPH code by \citet{GeretshauserEtal:preprint}.
However, as it was already pointed out in the laboratory section, the compressive strength relation has to be considered dynamically. The only
free parameter $p_{\rm m}$ (see Sect.\ \ref{sec:shift_strength_curve}) cannot be determined by experiments. Hence it has to be constrained by a
numerical parameter study. We will use the stopping time of the impacting glass bead as reference for this parameter.

In addition, a relation for the shear strength is also very hard to measure in the laboratory. Therefore, we suggest three simple relations
depending on the dynamic compressive strength and tensile strength relations and use the qualitative comparison of the filling factor profile
under the glass bead after impact to constrain this unknown quantity.

Finally, we utilize the remaining experimentally measured independent features of the experiments described in the laboratory section to
validate our calibration.

\subsection{\label{sec:benchmark-setup}Benchmark test - setup}

The given experimental setup (see Sect.\ \ref{sec:dyn_compr_exp}) was modeled with high resolution in two dimensions. Initially, the SPH
particles were put on a triangular grid. All simulations were performed with the influence of gravity taken into account.

The projectile was modelled with a circle of 1.1~mm in diameter consisting of 1519 SPH particles. Its material properties were simulated using
the Murnaghan equation of state
\begin{equation}
    p = \left( \frac{K_0}{n} \right) \left[ \left( \frac{\rho}{\rho_0} \right)^n - 1 \right]
\end{equation}
with $\rho_0 = 2540$~kg~m$^{-3}$ (total  2D mass per unit length $m_\mathrm{2D} = 2.4\cdot10^{-3}$~kg~m$^{-1}$), $K_0 = 5.0 \cdot 10^9$~Pa and
$n = 4$.  The density has been chosen such that it matches the experimental specifications. The other material parameters are similar to those
of sandstone. They can be found together with the Murnaghan equation of state in \citet{Melosh:1989}. The exact choice of the bulk modulus $K_0$
and the Murnaghan exponent $n$ does not have significant effects. The glass bead was treated as fully elastic.  The impact velocity was
0.65~m~s$^{-1}$.

The dust sample was modelled as a $8 \times 5$~mm$^2$ rectangle with 64421 SPH particles. About 0.15~mm at the bottom and 0.56~mm at each side
of the rectangle were used as reflecting boundary by setting their acceleration to zero at each time step. The porous material was simulated by
using the modified version of the Sirono model presented in Sect.\ \ref{sec:sph_basics}. The initial density was expressed via the filling
factor $\phi = \rho/\rho_0$ with $\phi = 0.15$ and $\rho_0 = 2000$~kg~m$^{-3}$. For the tensile strength we used the semianalytical relation,
derived in Sect.\ \ref{sec:tensile} (Eq.\ \ref{eq:tensile_strength}) that matches the findings of \citet{BlumSchraepler:2004} and paper I. The
bulk modulus was modeled by a power law
\begin{equation}
    K(\rho) = K_0 \left( \frac{\rho}{\rho_{\rm i}} \right)^4
\end{equation}
with $K_0 = 300$~kPa and the initial density of the dust aggregate $\rho_i = 300$~kg~m$^{-3}$. The bulk modulus $K_0=\rho_{\rm i} c^2$ for
uncompressed material was determined by the measurement of the sound speed, which is $c=30$~m~s$^{-1}$ \citep{BlumWurm:2008,PaszunDominik:2008}.

For the compressive strength, several different relations were tested. At first we adopted the relation from the uniaxial experiments by
\citet{BlumSchraepler:2004}. Secondly we used the omnidirectional compression curve presented in this paper. After it turned out that a modified
relation for the dynamical compressive strength curve had to be considered, the omnidirectional compression curve (Eq.\ \ref{eq:strength_curve})
was shifted towards lower pressure regimes using the  free parameter $p_{\rm m}$ (see Sect.\ \ref{sec:shift_strength_curve}).

Since no experimental data was available for the shear strength $Y$, parameter studies were carried out with three different relations
$Y(|T|,\Sigma)$: $Y = |T|$, $Y = \Sigma$ and, following \citet{Sirono:2004}, $Y = \sqrt{\Sigma |T|}$, which represents the geometric mean of
both quantities.

Due to reasons of stability, the two materials in contact (solid projectile, dusty target) have to be separated by artificial viscosity. We use
the approach by \citet{Monaghan:1983} and apply an $\alpha$-viscosity of $1.0$ to all particles of the sphere and all particles interacting with
the sphere. All other dust-sample particles were simulated without artificial viscosity following \citet{Sirono:2004}.  All details regarding
the SPH code can be found in \citet{GeretshauserEtal:preprint}.

\subsection{Calibration procedure}
\label{sec:calibration-procedure}

With the aim of reproducing the experimental results presented  in the laboratory section, an SPH simulation using the omnidirectional
compressive strength curve (ODC) was conducted. In the resulting pressure regime the ODC relation and the relation from
\citet{BlumSchraepler:2004} are almost identical. Therefore they can be treated as one case.

\begin{figure}
    \center
    \includegraphics[width=7.5cm,]{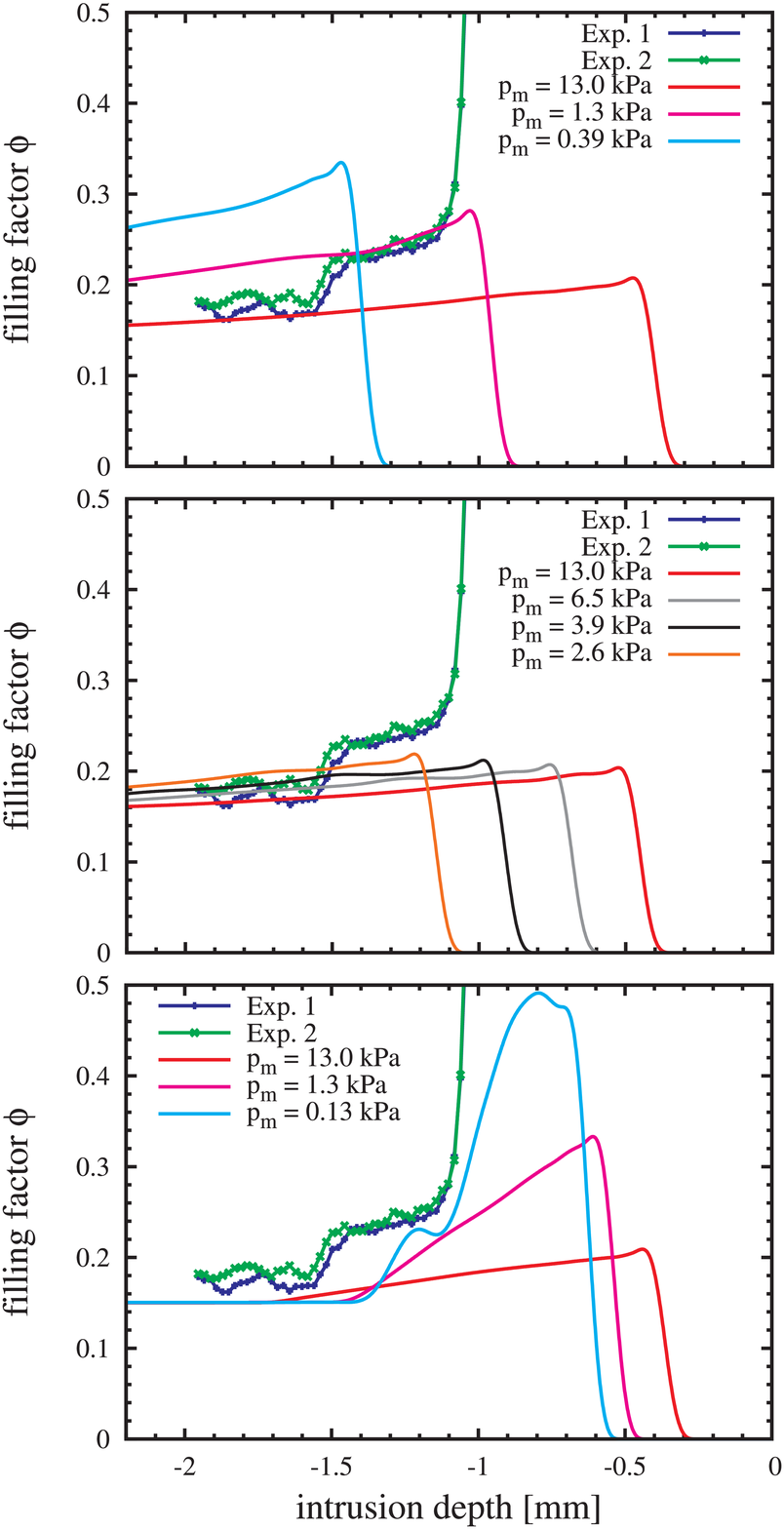}
    \caption{\label{fig:vert_dens}The filling factor measured along a line through the center of the sphere, perpendicular to the bottom of the dust
    sample (lines with blue and green crosses). The initial surface of the dust sample is situated at 0~mm and the steep slope at the
    right end of the experimental curves marks the transition from the dust sample to the glass sphere. The other curves are numerical simulations,
    varying the shear strength model and the material softness $p_\mathrm{m}$.  The shear strength relation was \textbf{(top)} $Y =
    \sqrt{\Sigma |T|}$, \textbf{(center)} $Y = \Sigma$, and \textbf{(bottom)} $Y = |T|$.}
\end{figure}

The impact velocity of the 1.1~mm glass bead was 0.65~m~s$^{-1}$ and we will compare the results of the simulation with the vertical density
profile along a line through the center of the sphere perpendicular to the bottom of the dust sample, which was measured with x-ray
micro-tomography as described in Sect. \ref{sec:dyn_compr_exp}. Figure \ref{fig:vert_dens} shows the results for three different shear strength
models, which are compared to two density profiles as measured in the experiments (lines with blue and green crosses). The initial surface of
the dust sample is at 0~mm.

For the original ODC relation ($p_\mathrm{m}=13$~kPa), the simulations for all shear strength models  resulted in a much too shallow intrusion
depth and an insufficient maximum filling factor underneath the sphere. These findings indicated, that the compressive strength curve had to be
modified in order to reproduce the experimental data. Therefore, we performed a parameter study varying the parameter $p_{\rm m}$, i.e.\
shifting the compressive strength curve to lower pressures for the different shear strength models. For the complete study see
\citet{GeretshauserEtal:preprint}. Independent experiments (paper III) also support a lower $p_{\rm m}$ which can quantitatively explain the
amount compression in bouncing collisions.

A significant increase of the intrusion depth was only observed in case of $Y = \sqrt{\Sigma |T|}$ and $Y = \Sigma$ (see
Fig.~\ref{fig:vert_dens}, top and center). In case of $Y = |T|$ the intrusion depth hardly changed with decreasing $p_{\rm m}$
(Fig.~\ref{fig:vert_dens}, bottom). Since the shear strength remained constant and changing $p_{\rm m}$ did not have a significant effect, it
can be concluded that shearing plays an important role during the intrusion.

Compared to the other cases, the shear strength reaches its highest values in the $Y = |T|$ case. Hence, the material can hardly be pushed away
due to shear and has to be compressed. Therefore, the highest filling factors can be found in this case (see Fig.~\ref{fig:vert_dens}, bottom).
The $Y = \Sigma$ model yields the lowest shear strength values. Hence, material is mostly sheared aside, less material is compressed and
therefore this model leads to filling factors below the reference data (see Fig.~\ref{fig:vert_dens}, center).

\begin{figure}[t]
    \center
    \includegraphics[angle=-90, width=7.5cm]{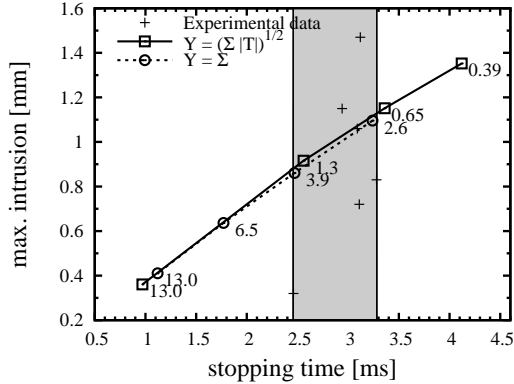}
    \caption{\label{fig:time-intrusion}Stopping time - intrusion diagram. Experimental data for spheres of 1~mm radius. Labels indicate the $p_{\rm
    m}$ values (in kPa) for the modification of the compressive strength curve. For the $Y = \Sigma$ model the best match in stopping time and
    intrusion depth is found for $p_{\rm m} = 3.9$~kPa. For the $Y = \sqrt{\Sigma |T|}$ model the best approximation is given for $p_{\rm m} =
    1.3$~kPa.}
\end{figure}

Figure \ref{fig:time-intrusion} shows intrusion depth over  stopping  time regarding the shear strength models $Y = \Sigma$ and $Y =
\sqrt{\Sigma |T|}$ for all $p_{\rm m}$. Since $Y = |T|$ did only yield insufficient intrusion depths, this model was omitted here. A good
time/depth match was achieved for $p_{\rm m} = 3.9$~kPa using $Y = \Sigma$ and for $p_{\rm m} = 1.3$~kPa using $Y = \sqrt{\Sigma |T|}$. However,
the $Y = \Sigma$ model cannot reproduce the high values in the vertical filling factor profile (Fig.~\ref{fig:vert_dens}, center) whereas the $Y
= \sqrt{\Sigma |T|}$ model yields an almost perfect match (Fig.~\ref{fig:vert_dens}, top). Therefore, the latter with $p_{\rm m} = 1.3$~kPa
gives a good match in Fig.~\ref{fig:vert_dens} (top) as well as in Fig.~\ref{fig:time-intrusion} and is therefore used for further simulations.
A more detailed study on the determination on the $p_\mathrm{m}$ value can be found in \citet{GeretshauserEtal:preprint}.

Hereby, we have determined parameters for all previously unknown material relations and thus have calibrated the SPH model with respect to the
presented experiments.
\begin{figure}[t]
    \center
    \includegraphics[height=7.5cm,angle=90]{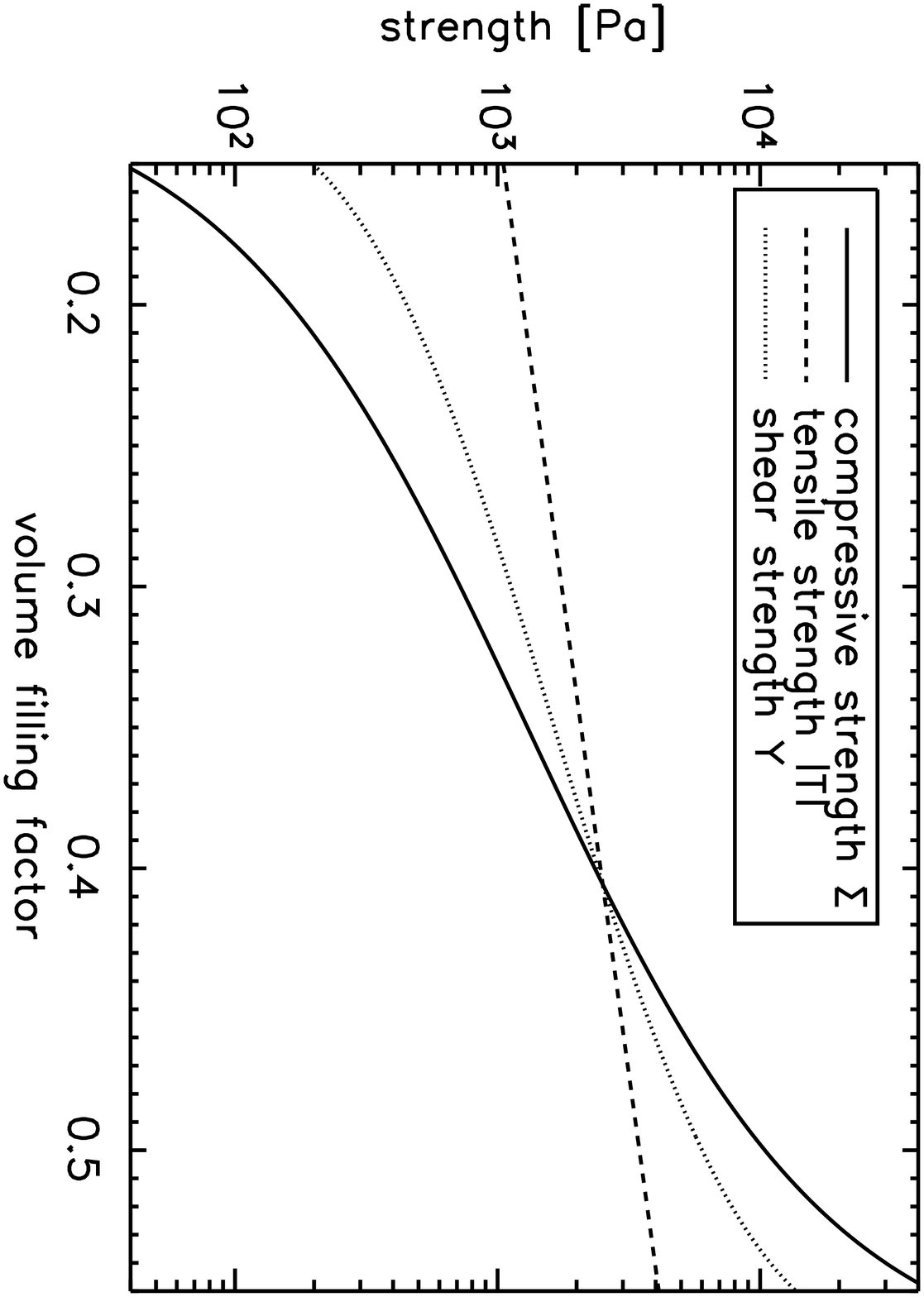}
    \caption{\label{fig:strength_curves}Compressive strength curve $\Sigma(\phi)$ (Eq.\ \ref{eq:strength_curve}) for $p_{\rm m}=1.3$~kPa, tensile
    strength $|T|$ (Eq.\ \ref{eq:tensile_strength}), and shear strength $Y=\sqrt{\Sigma\cdot|T|}$.}
    \includegraphics[angle=-90, width=7.5cm]{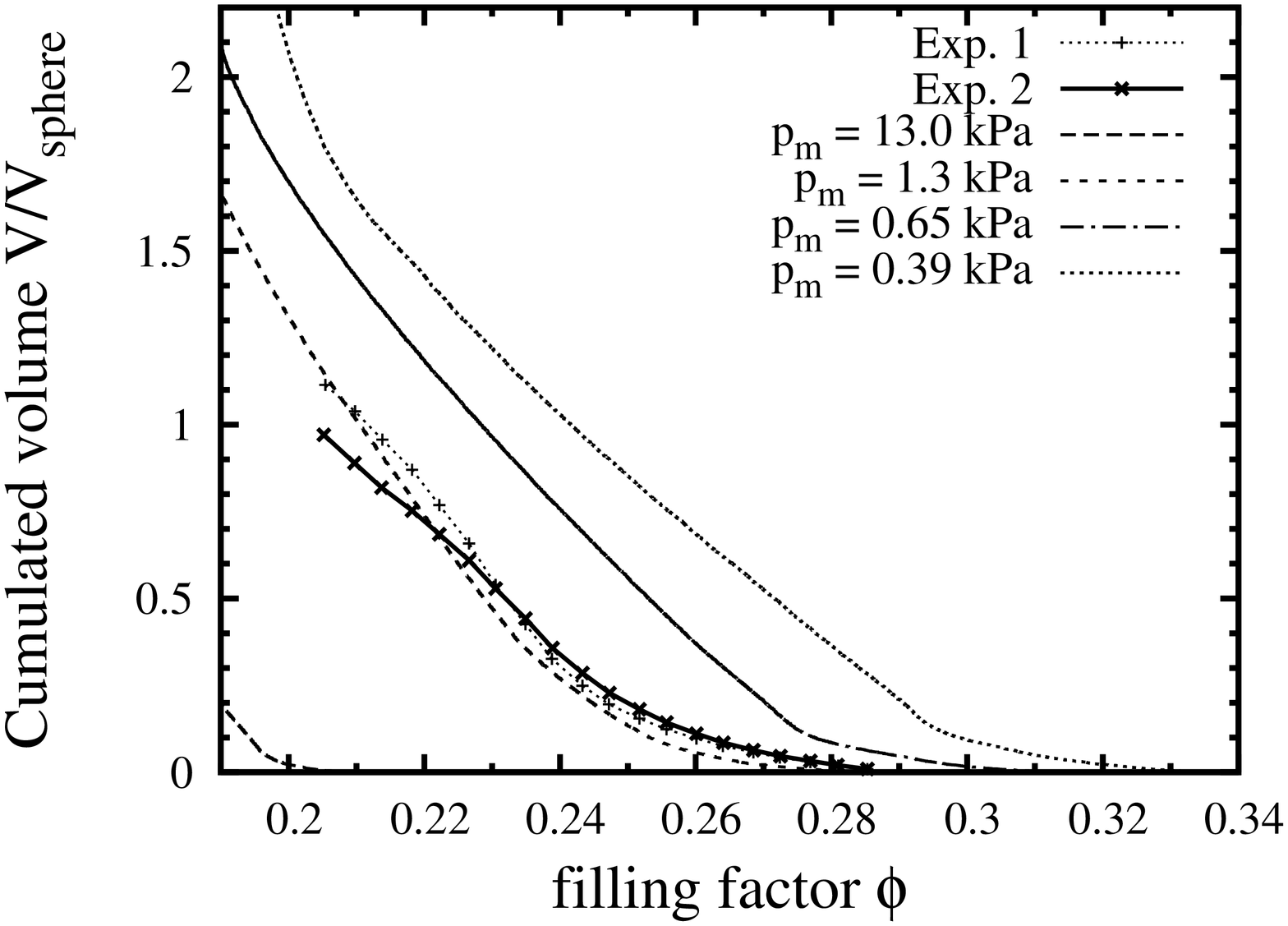}
    \caption{\label{fig:mass_YSir}Cumulated volume over filling factor. While $p_{\rm m} = 13$~kPa, i.e.\ the unmodified ODC relation, and $p_{\rm
    m} = 0.39$~kPa as well as $p_{\rm m} = 0.65$~kPa yield too small and too high compression values, respectively, $p_{\rm m} = 1.3$~kPa matches
    very well for $\phi \gtrsim 0.22$. The experimental data are identical to those shown in Fig.~\ref{fig:vol_fillfactor} (bottom)}
\end{figure}
The resulting strength curves of compression (Eq. \ref{eq:strength_curve}, $p_{\rm m} = 1.3$~kPa), tension (Eq.\ \ref{eq:tensile_strength}) and
shear ($Y = \sqrt{\Sigma |T|}$) are illustrated in Fig.~\ref{fig:strength_curves}.

However, the fact that the filling factor does not rapidly drop to $\sim 0.15$ at a depth of 1.5~mm requires further investigation.

\subsection{Reproducing Experimental Features}

Since intrusion time and depth as well as the filling factor profile underneath the sphere have been used to determine $p_{\rm m}$ and the
correct shear strength model, further features have to be reproduced in order to validate the calibration.

\begin{figure}[t]
    \center
    \includegraphics[width=5cm,angle=90]{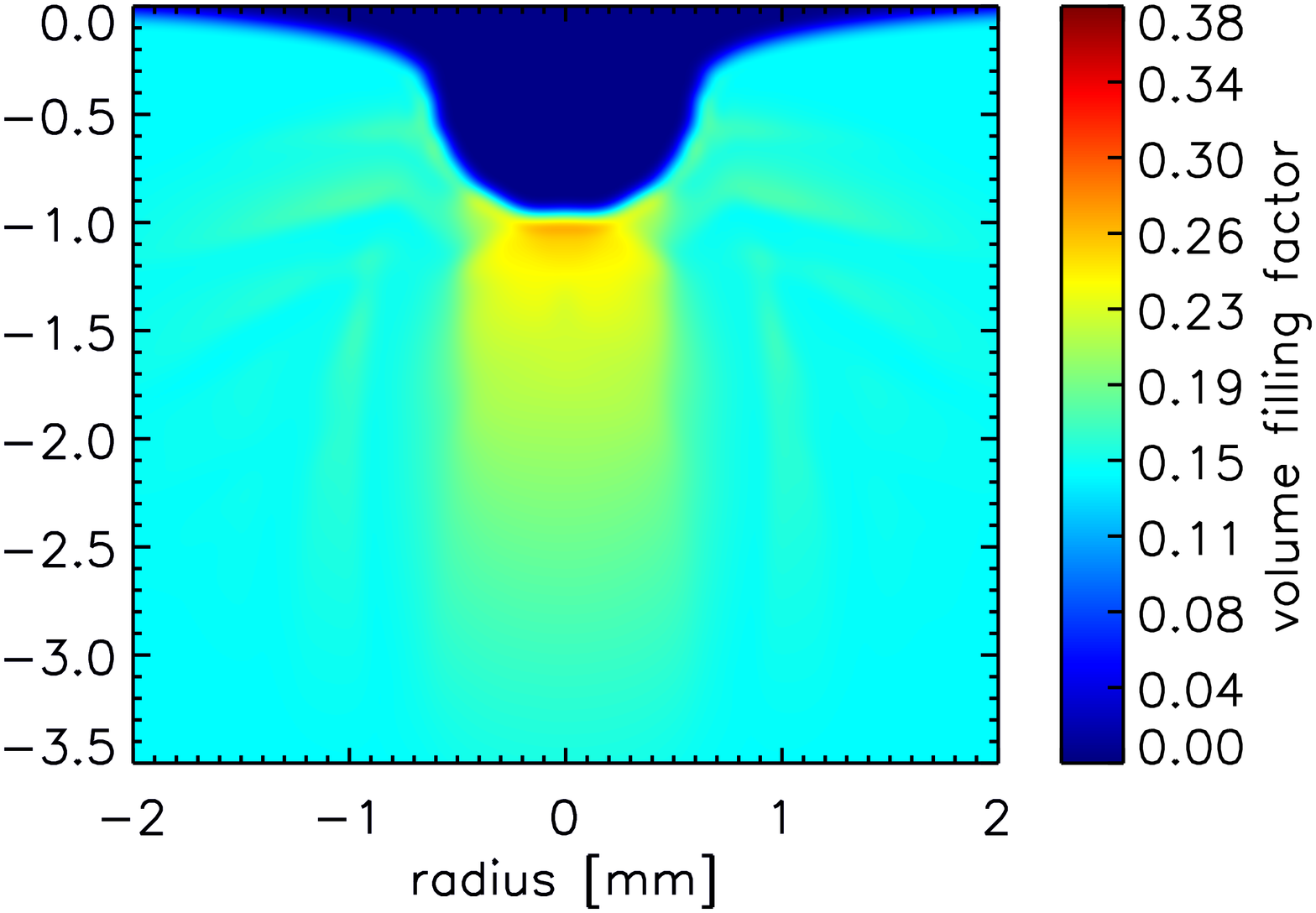}
    \caption{\label{fig:color_contour_sph}Spatially distributed compression as calculated in the SPH simulation with $Y=\sqrt{\Sigma\cdot|T|}$ and
    $p_{\rm m} = 1.3$~kPa; same color scale as Fig.~\ref{fig:color_contour}; the projectile is not plotted. Although the filling factor of
    compressed material is comparable to the one in the experiments, the compressed volume reaches significantly deeper.}
    \includegraphics[angle=-90, width=7.5cm]{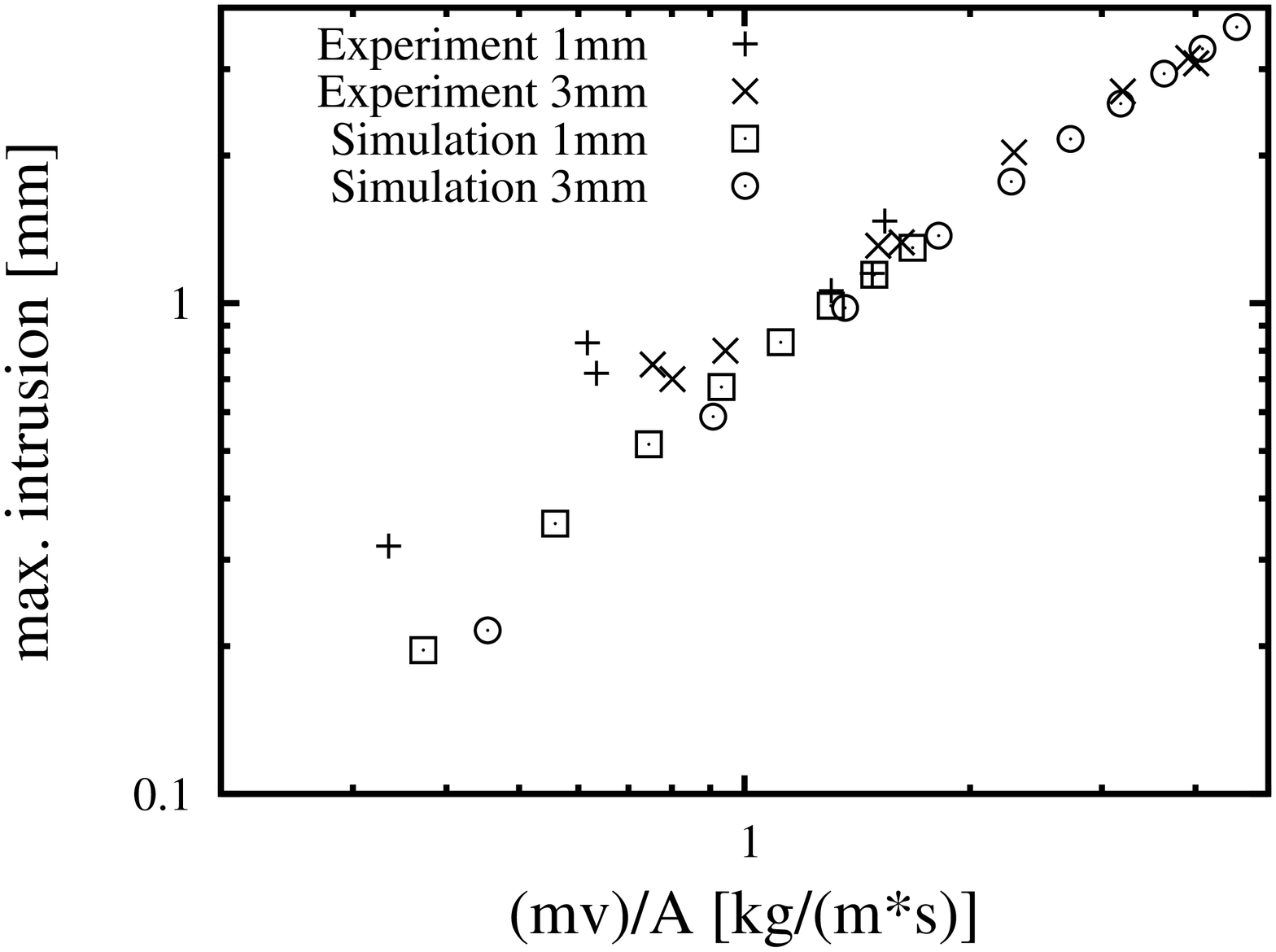}
    \caption{\label{fig:momentum-intrusion}In the momentum-intrusion relation, the agreement between simulation and experimental results is very
    good for values of $mvA^{-1}\gtrsim1$~kg~m$^{-1}$~s$^{-1}$.}
\end{figure}

One of these features is the cumulated volume over filling factor relation (Fig.~\ref{fig:mass_YSir}). While the filling factor profile only
displays a cut through the compressed volume, this curve represents the total compressed volume with its filling factors. Both curves are not
fully, but mostly independent from each other. The chosen model and $p_{\rm m}$ value yield an almost perfect match for filling factors $>0.22$.
The deviation for lower filling factors is due to the larger amount of compressed volume. This effect was already seen in the filling factor
profile and is also very prominent in the comparison of the spatially density distribution plots (compare Figs. \ref{fig:color_contour} and
\ref{fig:color_contour_sph}).

Another feature to be reproduced is the relation $D\propto mvA^{-1}$ found in a similar way in the drop experiments (cf.
Fig.~\ref{fig:penetration_depth}). We performed a series of 2D simulations with spheres of 1 mm and 3 mm diameter and evaluated the maximum
intrusion depth with respect to the impact velocity $v$. The latter was varied from 0.1~m~s$^{-1}$ to 1.0~m~s$^{-1}$ in steps of 0.1~m~s$^{-1}$.

\begin{figure}[t]
    \center
    \includegraphics[height=7.5cm,angle=90]{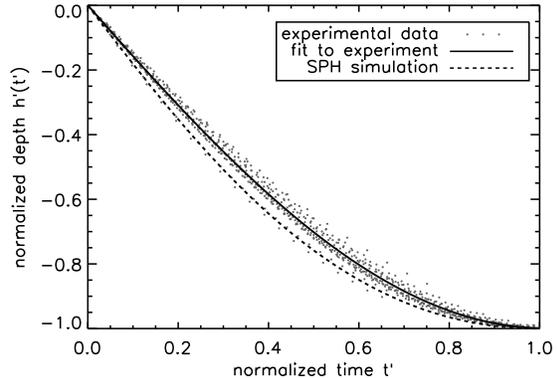}
    \caption{\label{fig:norm_dec_curve}Normalized deceleration curve compared to the results. The deceleration curve in the SPH simulation is
    slightly lower than the experimentally observed sine curve, but well within the errors. This effect will be be analyzed in future work. However,
    the range of experimental data encompasses the simulation results.}
\end{figure}

2D simulation and experiment cannot be compared directly due to the different geometry (the 2D setup represents a cut through an infinitely long
cylinder).  The advantage of using the quantity $mvA^{-1}$ instead of the more accurate Eq.\ \ref{eq:penetr_depth} is given by the fact that the
former can be ``converted'' into 2D by the following correction:
\begin{equation}
    \frac{m_{3D}v}{A_{3D}} = \frac{\frac{4}{3} \pi r^3 \rho \cdot
    v}{\pi r^2} = \frac{8}{3\pi} \, \frac{\pi r^2 \rho \cdot v}{2r}
    = \frac{8}{3\pi} \, \frac{m_{2D}v}{A_{2D}}
\end{equation}
In comparison with the experimental results, the data from the simulation matches very well for $mvA^{-1} > 1.0$~kg~m$^{-1}$~s$^{-1}$
(Fig.~\ref{fig:momentum-intrusion}). For smaller values the simulation yields a shallower intrusion than the reference experiments, which,
however, also show significant scattering in this range.

Comparing the simulated and experimentally acquired normalized deceleration curves (Fig.~\ref{fig:norm_dec_curve}), the simulated data slightly
deviate from the experimental mean but remain within standard derivation limits. The deviation could arise from the geometric difference of the
2D and 3D case and has to be investigated in future works.


\section{\label{sec:application}Application of SPH to Dust Collisions in PPDs, Conclusions, and Outlook}
In this section we will present some preliminary applications of SPH simulations to dust collisions in protoplanetary disks. We will present
two examples of previously unfeasible calculations of inter-particle collisions among macroscopic dust aggregates and will qualitatively compare
them to similar dust experiments performed in the  laboratory. Then, we will speculate about how the SPH code should be used in research on
protoplanetary growth. Finally we will sketch future work in preparation.

\subsection{\label{sec:comparison}Qualitative comparison between SPH simulations and laboratory experiments}
The strength of the SPH simulations -- besides the well-known examples in hyper-velocity collisions -- over laboratory experiments and
molecular-dynamics simulations is that low-velocity collisions among arbitrary dust aggregates can be investigated. Here, we show two examples
recently observed in the lab, which can so far not be described by any other model. Example 1 deals with the frequently-observed bouncing
collisions in aggregate-aggregate interactions. Example 2 describes the impact of a single dust aggregate onto a solid flat target, which shows
the co-occurrence of (partial) sticking {\em and} fragmentation.

\subsubsection{\label{sec:example1}Example 1}
\begin{figure}[t]
    \center
    \includegraphics[width=7.5cm]{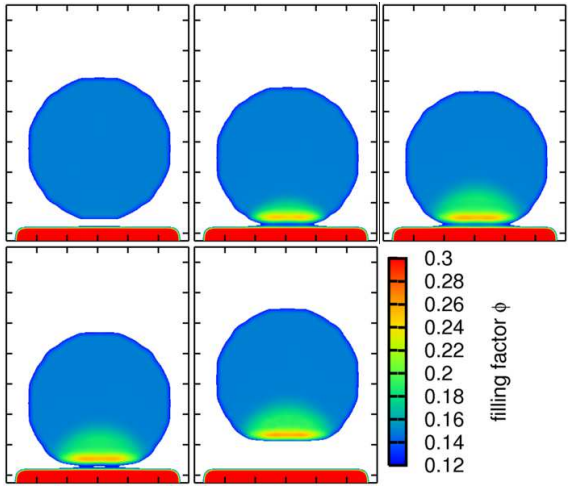}
    \caption{\label{fig:bouncing}Sequence of snapshots of an SPH simulation of a fluffy dust aggregate with a radius of 0.5 mm, impacting a solid
    target at a velocity of 0.2 $\rm m\,s^{-1}$. The time differences between subsequent images are 0.35 ms, 0.32 ms, 0.23 ms, and 4.45 ms,
    respectively. The colors denote different degrees of internal compaction. \emph{[See the electronic edition of the Journal for accompanying mpeg
    animations.]}}
\end{figure}
Bouncing in collisions between dust aggregates has been observed in many laboratory experiments \citep[][paper III; Hei{\ss}elmann et al., in
prep., will appear in this series]{BlumMuench:1993,LangkowskiEtal:2008}, although molecular-dynamics simulations always show a direct transition
from sticking to fragmentation when the collision energy exceeds a threshold value \citep{DominikTielens:1997,WadaEtal:2007,WadaEtal:2008}.
Nature obviously chooses a wider bouncing transition between those two stages, at least for aggregates above a certain size. It turns out that
the SPH method is capable of describing the bouncing phase quite well. We have run a 3D SPH simulation of a low-velocity impact of a 1~mm
(diameter) fluffy aggregate onto a flat target. Due to symmetry arguments, this is identical to a two-aggregate (central) collision with twice
the collision velocity. In our case, the aggregate was composed of 33,377 SPH particles and had an initial volume filling factor of 0.15. All
other material parameters were identical to those in the previous section, i.e.\ $K_0 = 300$~kPa, $p_{\rm m}=1.3$~kPa, $Y = \sqrt{\Sigma |T|}$.
The impact velocity was 0.2 $\rm m\,s^{-1}$, matching exactly the situation in the aggregate-wall experiments performed in paper III and also
those in the aggregate-aggregate collisions investigated by Hei{\ss}elmann et al.\ (in prep.) with a collision speed of 0.4 $\rm m\,s^{-1}$.
Fig.~\ref{fig:bouncing} shows a sequence of snapshots with a cut through the center of the aggregate, indicating the internal compaction due to
the impact. Our simulation can correctly predict the coefficient of restitution of $\sim 0.2$ \citep[][Hei{\ss}elmann et al., in
prep.]{BlumMuench:1993}, although details in the compaction behavior still deviate from the laboratory results, which might be caused by
insufficient resolution in the SPH simulation.

\subsubsection{\label{sec:example2}Example 2}
\begin{figure}[t]
    \center
    \includegraphics[width=7.5cm]{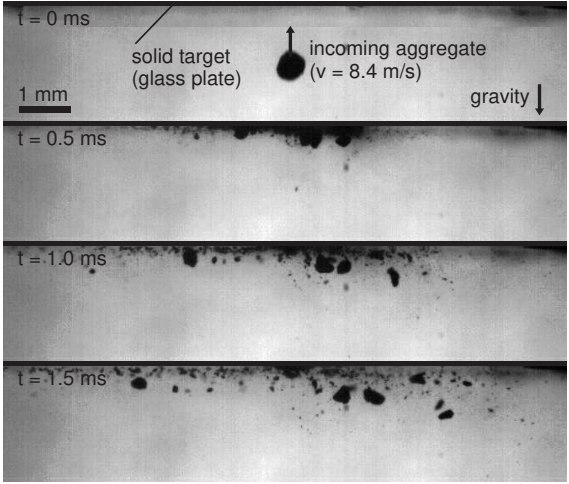}
    \caption{\label{fig:stfrlab}Image sequence of an experiment, in which a fluffy dust aggregate impacts a solid target at 8.4~m~s$^{-1}$. Part of
    the aggregate sticks to the target after the collision. \emph{[See the electronic edition of the Journal for accompanying mpeg animations.]}}
\end{figure}
\begin{figure}[t]
    \center
    \includegraphics[width=7.5cm]{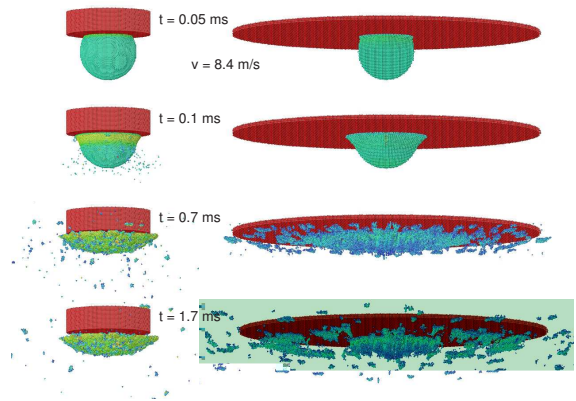}
    \caption{\label{fig:stfrsph}Image sequence of an SPH simulation, identical to the experiment shown in Fig.~\ref{fig:stfrlab}. The simulation
    with the calibrated parameters \textbf{(left)} cannot reproduce the experimental results, whereas a simulation using an unidirectional compression
    curve and $\Sigma=|T|$ \textbf{(right)} can reproduce the qualitative findings of Fig. \ref{fig:stfrlab}. \emph{[See the electronic edition of
    the Journal for accompanying mpeg animations.]}}
\end{figure}
In the previous example, we have seen that bouncing marks the broad transition regime between sticking and fragmentation. However, in the case
of the impact of a dust aggregate onto a solid target, laboratory experiments have shown that, for impact experiments above the fragmentation
threshold, fragmentation is always accompanied by partial sticking of the aggregate to the target. This effect was first found by
\citet{WurmEtal:2005} for compacted dust aggregates and impact velocities above 25 $\rm m\,s^{-1}$ and later confirmed in our laboratory for
$\phi=0.35$ aggregates and impact velocities above 1 $\rm m\,s^{-1}$. Fig.~\ref{fig:stfrlab} shows an image sequence of an impact experiment
with fragmentation and partial sticking. An average of 10~\% of the projectile mass sticks to an initially smooth target at normal impact, which
is consistent with the low velocity results of \citet{WurmEtal:2005}. The remainder of the projectile mass is fragmented into a power-law mass
distribution \citep[see][]{BlumMuench:1993}. The fragments leave the target under extremely flat angles. Our SPH simulation
(Fig.~\ref{fig:stfrsph}, left) featuring the calibration parameters of Sect. \ref{sec:calibration-procedure} cannot reproduce the fragmentation
behavior seen in the experiments. Here, the predominant part of the dust sample sticks to the target. Only a few bigger chunks and single SPH
particles burst off. However, a simulation with the same setup, but using the shifted unidirectional compressive strength relation (Sect.
\ref{sec:compr_curve}) and a shear strength that is equal to the tensile strength, matches the experimental observations at least qualitatively
(Fig.~\ref{fig:stfrsph}, right). From that we conclude that the SPH code is in principle capable of simulating fragmentation of highly-porous
aggregates, even without the damage model adopted in the original \citet{Sirono:2004} porosity model.

We conclude that the shear model $Y=\Sigma^{0.5}\cdot |T|^{0.5}$ tested for the dynamic compression experiments (Sect.
\ref{sec:benchmark-setup}) is unable to explain the fragmentation findings which are rather dominated by shear and tension, whereas a shear
model $Y=\Sigma^{0}\cdot |T|^{1}$ shows qualitative agreement. The imperfect shear model  can also be  responsible for the narrow but deep
compressed volume in Fig. \ref{fig:color_contour_sph} compared to Fig. \ref{fig:color_contour}. A future task will therefore be to refine the
shear calibration in a way that we will use $Y=C\cdot \Sigma^{\alpha}\cdot |T|^{1-\alpha}$ with the free parameters $C$ and $\alpha$. Comprising
both experiments for calibration we will be able to find a shear model that can reproduce both cases.

\subsection{\label{sec:use}Use of the SPH code in research on protoplanetary growth}
The above examples show that the SPH method is a powerful tool to investigate the outcomes of protoplanetary dust collisions. When properly
calibrated with laboratory experiments, SPH calculations allow access to parameter-space regions that are unavailable to laboratory experiments.
Whereas molecular-dynamics simulations can be used for studying collisions of very small dust aggregates, SPH is most useful for very large
samples. Such samples, particularly those with fluffy compositions, cannot be built or treated in laboratories, and the experimental study of
collisions seems impossible.

A particularly interesting and still unsolved problem is the dichotomy in the collision behavior of pairs of dust aggregates with similar and
different sizes, respectively. In paper II we found sticking by deep penetration for impacts of mm-sized dusty projectiles into flat, cm-sized
dusty targets (``projectile-target'' collisions) above $\sim 1\, \rm m\, s^{-1}$. Both dust aggregates, projectile and target, consisted of
identical particles and had equal porosity. Using similar dust aggregates, but giving projectile and target comparable size
(``projectile-projectile'' collisions), \citet{BlumMuench:1993} and Hei{\ss}elmann et al.\ (in prep.) found that collisions either lead to
bouncing or to fragmentation. Bouncing instead of sticking was also observed in paper II when the target aggregates were prepared such that the
local radius of curvature corresponded to the projectile's radius. To find out where the boundary between ``projectile-target'' and
``projectile-projectile'' collisions occurs, will be one of our future applications of our SPH code.

\subsection{\label{sec:future}Future work}
We have only begun to explore the potentials of SPH simulations of collisions between protoplanetary dust aggregates. Before we can start to
investigate the full parameter space in protoplanetary dust collisions, i.e. before we can begin to find out what the collisional outcome is for
all combinations of aggregate size, porosity, collision velocity, impact angle, state of rotation, temperature and state of sintering, material
and size (distribution) of the constituent dust grains, etc., the material parameters of macroscopic dust aggregates have to be fully explored.
This will be the next task in our investigation. To achieve this, we will perform more calibration experiments of the type described in this
paper for dust aggregates of various compositions and porosities. In addition to that, other calibration experiments will be explored, like the
ones described in Sects.\ \ref{sec:example1} and \ref{sec:example2}.


\subsection*{Acknowledgements}
We thank M.-B. Kallenrode and the University of Osnabr{\"u}ck for providing access to the XRT setup  and Jens Teiser for the first feasibility
tests for the experiments. The SPH simulations were performed on the university and bwGriD clusters of the computing center (ZDV) of the
University of T{\"u}bingen. This project was funded by the Deutsche Forschungsgemeinschaft within the Forschergruppe 759 ``The Formation of
Planets: The Critical First Growth Phase'' under grants Bl 298/7-1, Bl 298/8-1, and Kl 650/8-1.

\begin{footnotesize}
    \bibliography{../../../../literatur}
\end{footnotesize}

\end{document}